\documentclass[12pt]{article}
\pdfoutput=1

\usepackage[height=23truecm,width=16.4truecm,top=2.2truecm]{geometry}

\usepackage[skip=6pt,indent]{parskip}

\usepackage[utf8]{inputenc}
\usepackage{graphicx}
\usepackage{amsmath}
\usepackage{amssymb}
\usepackage{hyperref}
\hypersetup{colorlinks,citecolor=TokiwaIro,linkcolor=RuriIro,urlcolor=RuriIro,linktocpage}
\usepackage{cite}
\usepackage{braket}
\usepackage{bm}
\usepackage{bbm}
\usepackage{color}
\usepackage[svgnames,table]{xcolor}
\usepackage{float}
\usepackage{hhline}
\usepackage{multirow}
\usepackage{indentfirst}
\usepackage{latexsym}
\usepackage{mathtools}
\usepackage{cancel}
\usepackage{slashed}
\usepackage{colortbl}
\usepackage[mathscr]{euscript}
\usepackage{enumitem}
\usepackage{subcaption}
\usepackage{comment}
\usepackage{amsthm}

\numberwithin{equation}{section}

\allowdisplaybreaks

\definecolor{RuriIro}{rgb}{0.,0.28,0.60}
\definecolor{TokiwaIro}{rgb}{0.,0.39,0.16}

\definecolor{dred}{rgb}{0.7,0.15,0.09}
\definecolor{dblue}{rgb}{0,0.12,0.64}
\definecolor{dgreen}{rgb}{0.2,0.51,0.19}
\definecolor{pegn}{rgb}{0.33,0.51,0.14}

\newcommand{\nn}{\nonumber}

\newcommand{\mc}{\mathcal}
\newcommand{\mr}{\mathrm}

\newcommand{\mbb}{\mathbb}

\newcommand{\del}{\partial}
\newcommand{\ol}{\overline}
\newcommand{\dd}{\mathrm{d}}
\newcommand{\ee}{\mathrm{e}}
\newcommand{\iu}{\mathrm{i}}

\renewcommand{\Re}{\mathop{\mathrm{Re}}}
\renewcommand{\Im}{\mathop{\mathrm{Im}}}

\newcommand{\SL}{\mathrm{SL}}
\newcommand{\PSL}{\mathrm{PSL}}
\newcommand{\rU}{\mathrm{U}}

\begin{document}

\begin{titlepage}

\begin{flushright}
{\ttfamily
EPHOU-24-005
}
\end{flushright}

\vspace{1cm}

\begin{center}

{\Large \bfseries
Moduli stabilization 
in finite modular symmetric models
}

\vspace{1cm}

\renewcommand{\thefootnote}{\fnsymbol{footnote}}
{%
\hypersetup{linkcolor=black}
Yoshihiko Abe$^{1}$\footnote[1]{yabe3@wisc.edu},
\ 
Komei Goto$^{2}$\footnote[2]{komeigt@keio.jp},
\ 
Tetsutaro Higaki$^{2}$\footnote[3]{thigaki@rk.phys.keio.ac.jp},
\\
Tatsuo Kobayashi$^{3}$\footnote[4]{kobayashi@particle.sci.hokudai.ac.jp},
\ and
Kaito Nasu$^{3}$\footnote[5]{k-nasu@particle.sci.hokudai.ac.jp}
}%
\vspace{8mm}

{\itshape%
$^1${Department of Physics, University of Wisconsin-Madison, Madison, WI 53706, USA}\\
$^2${Department of Physics, Keio University, Yokohama 223-8533, Japan}\\
$^3${Department of Physics, Hokkaido University, Sapporo 060-0810, Japan}
}%

\vspace{8mm}

\end{center}

\abstract{
We study vacua of moduli potential consisting of multiple contribution of modular forms in a finite modular symmetry.
If the potential is given by a single modular form, the Minkowski vacuum is realized at the fixed point of the modular symmetry.
We show that the de Sitter vacuum is realized with a multiple modular form case and obtain a non-trivial vacuum which is away from the fixed point, i.e. a large modulus vacuum expectation value, depending on the choice of the weight and representation of the modular forms.
We study these vacua numerically and analytically.
It is also found that the vacua obtained in this paper preserve CP symmetry.
}

\end{titlepage}

\renewcommand{\thefootnote}{\arabic{footnote}}
\setcounter{footnote}{0}
\setcounter{page}{1}

\tableofcontents

\section{Introduction}

A modular symmetry is well-motivated from the geometrical symmetry of the higher dimensional theories such as superstring theory.
For example, the modulus parameter $\tau$ is the complex structure modulus when we consider a torus or its orbifold compactification.
Then $\tau$ is called the modulus field, which is a dynamical degree of freedom in the four-dimensional effective field theory.

The moduli fields play important roles both in particles physics and cosmology. These days, the flavor symmetry based on the modular group~\cite{Feruglio:2017spp} has been attracting much attention.
In the context of flavor physics, the three generations of quarks and leptons transform non-trivially under the modular flavor symmetry.
In addition, Yukawa couplings are assumed to be modular forms which are holomorphic functions of a modulus $\tau$ and also transform non-trivially as similar to the matter fields.
As discussed in Ref.~\cite{deAdelhartToorop:2011re}, we note that the (in)homogeneous finite modular group $\Gamma_N$ with the level $N \leq 5$ is isomorphic to the well-known permutation groups, such as $S_3$, $A_4$, $S_4$, and $A_5$.
The flavor physics based on these new flavor structures has been actively discussed in the literature~\cite{Feruglio:2017spp,Kobayashi:2018vbk,Penedo:2018nmg,Novichkov:2018nkm,Ding:2019xna,Liu:2019khw,Novichkov:2020eep,Liu:2020akv,Liu:2020msy,Kobayashi:2023zzc,Ding:2023htn}.

The scalar potential with a modular symmetry can generally be written with modular functions, and the modulus stabilization in modular flavor symmetric models is discussed in Refs.~\cite{Kobayashi:2019uyt,Kobayashi:2019xvz,Novichkov:2022wvg}. 
Such potential has also been studied in the context of the inflation models \cite{Kobayashi:2016mzg,Higaki:2015kta,Schimmrigk:2016bde,Lynker:2019joa,Schimmrigk:2021tlv,Abe:2023ylh,Ding:2024neh,King:2024ssx}.
Independently of the inflation model, the potential typically has the following form
\begin{align}
    V \sim \sum_{i=1} c_i (2\Im\tau)^{k_i}|Y_i(\tau)|^2,
    \label{eq:potential-demo}
\end{align}
where $c_i$ is a constant, $Y_i(\tau)$ denotes a modular form,
$k_i$ is a weight of $Y_i(\tau)$, and $i$ labels each modular form contribution.
When the potential contains only a single modular form $Y(\tau)$, i.e. $V = (2\Im\tau)^{k}|Y(\tau)|^2$, the vacuum is obviously given by a zero point of the modular form $Y(\tau_*)= 0$ if $Y=0$ exists.
Here, $\tau_*$ is a value at $Y=0$.
If there exist multiple contributions of the modular forms in the scalar potential as in Eq.~(\ref{eq:potential-demo}), the competition among $Y_i$ gives a vacuum away from $\tau_*$.
Such a new vacuum can realize a flavor hierarchy with a large vacuum expectation value (VEV) of $\tau$. On top of that, if a CP-breaking vacuum could be realized, the matter-anti-matter asymmetry could be explained.

In this work, we introduce matter fields coupled to modular forms and find that Eq.~(\ref{eq:potential-demo}) is realized.
Then, we study vacua of scalar potential consisting of one or two modular forms as in Eq.~\eqref{eq:potential-demo} with the modular symmetry and an (approximate) $R$-symmetry. 
Models with two modular forms help to understand how a stable vacuum is generated and whether or not the CP symmetry is broken.
For simplicity, we focus on the $A_4$ modular symmetry, which is a well-studied non-Abelian discrete flavor symmetry, and the scalar potential written with its singlet modular forms, $Y^{(4)}_{\bm{1}^{(\prime)}}$ and $Y^{(6)}_{\bm{1}}$.

This paper is organized as follows.
In Sec.~\ref{sec:modula-symmetry}, we give a brief review of modular symmetry and modular forms.
We also discuss the systematic expansion of the modular forms at three fixed points of the modular symmetry.
Our model is introduced in Sec.~\ref{sec:Model}. 
We show the details of the K\"{a}hler potential and superpotential, and give an overview of our models.
In Sec.~\ref{sec:single-modular-form-potential}, we show the potential profile with a single modular form
and find that a vacuum exists at the fixed point.
Sec.~\ref{sec:double-modular-forms-potential} shows how the stable vacuum at the fixed point shifts by an additional modular form contribution in numerical and analytically perturbative ways.
Sec.~\ref{sec:conclusion} is devoted to our conclusion.
The details of the modular forms and formulae used in our analysis are summarized in the appendices.

In the following parts, we will set the reduced Planck scale $M_P \approx 2.4 \times 10^{18}~\mr{GeV}$ to be unity otherwise stated.

\section{Modular symmetry and modular forms}
\label{sec:modula-symmetry}

In this section, we give a brief review of the modular symmetry and modular forms.
We also discuss the expansion of the modular forms around their fixed points.

\subsection{Modular symmetry}

\subsubsection{Modular symmetry and modular groups}

The homogeneous modular group $\Gamma \coloneqq \SL(2,\mbb{Z})$ is defined by 
\begin{align}
    \Gamma = \left\{
        \begin{pmatrix}
            a & b \\
            c & d
        \end{pmatrix} \bigg|\ 
        a, b, c, d \in \mbb{Z},
        \quad 
        ad - bc =1
    \right\},
\end{align}
which is generated by
\begin{align}
    S = \begin{pmatrix}
        0 & 1 \\
        -1 & 0
    \end{pmatrix},
    \qquad 
    T = \begin{pmatrix}
        1 & 1 \\
        0 & 1
    \end{pmatrix},
    \qquad 
    R = \begin{pmatrix}
        -1 & 0 \\
        0 & -1
    \end{pmatrix}.
\end{align}
These generators satisfy the following relations:
\begin{align}
    S^2 = R,
    \qquad 
    (ST)^3 = R^2 = S^4 = \bm{1},
    \qquad 
    TR = RT.
    \label{eq:STR-relations}
\end{align}
Under this $\SL(2,\mbb{Z})$ transformation, the modulus $\tau$ transforms as 
\begin{align}
    \gamma: \tau \mapsto \frac{a \tau + b}{c \tau + d},
    \qquad 
    \gamma \in \SL(2, \mbb{Z}).
    \label{eq:gamma-tau}
\end{align}
In particular, the generators, $S$ and $T$, act $\tau$ as 
\begin{align}
    S: \tau \mapsto - \frac{1}{\tau},
    \qquad 
    T: \tau \mapsto \tau + 1,
\end{align}
and $\tau$ is invariant under $R = - \bm{1}$.
This is called the modular transformation, and we introduce the modular group by $\bar{\Gamma} = \PSL(2, \mbb{Z}) = \SL(2, \mbb{Z})/ \mbb{Z}^R_2$, where $\mbb{Z}^R_2$ is generated by $R$.

The congruence subgroup of the level $N$, denoted by $\Gamma(N)$, is defined by
\begin{align}
    \Gamma(N) \coloneqq \left\{
        \begin{pmatrix}
            a & b \\
            c & d
        \end{pmatrix} \in \SL(2, \mbb{Z}),
        \ 
        \begin{pmatrix}
            a & b \\
            c & d
        \end{pmatrix} \equiv \begin{pmatrix}
            1 & 0 \\
            0 & 1
        \end{pmatrix}
        \mod N
    \right\}.
\end{align}
The quotients $\Gamma_N \coloneqq \bar{\Gamma} / \Gamma(N)$ for $N= 2, 3, 4,$ and $5$ are isomorphic to $S_3$, $A_4$, $S_4$, and $A_5$, respectively~\cite{deAdelhartToorop:2011re}.
In these quotients, $T$ also satisfies $T^N=\bm{1}$ with $N\leq 5$\footnote{
If $N > 5$, additional relations are required in order to close the algebra~\cite{deAdelhartToorop:2011re}.} , which generates an additional $\mbb{Z}^T_N$ symmetry.

\subsubsection{Fundamental domains and fixed points}

\begin{figure}[t]
    \centering
    \includegraphics[width=0.5\textwidth]{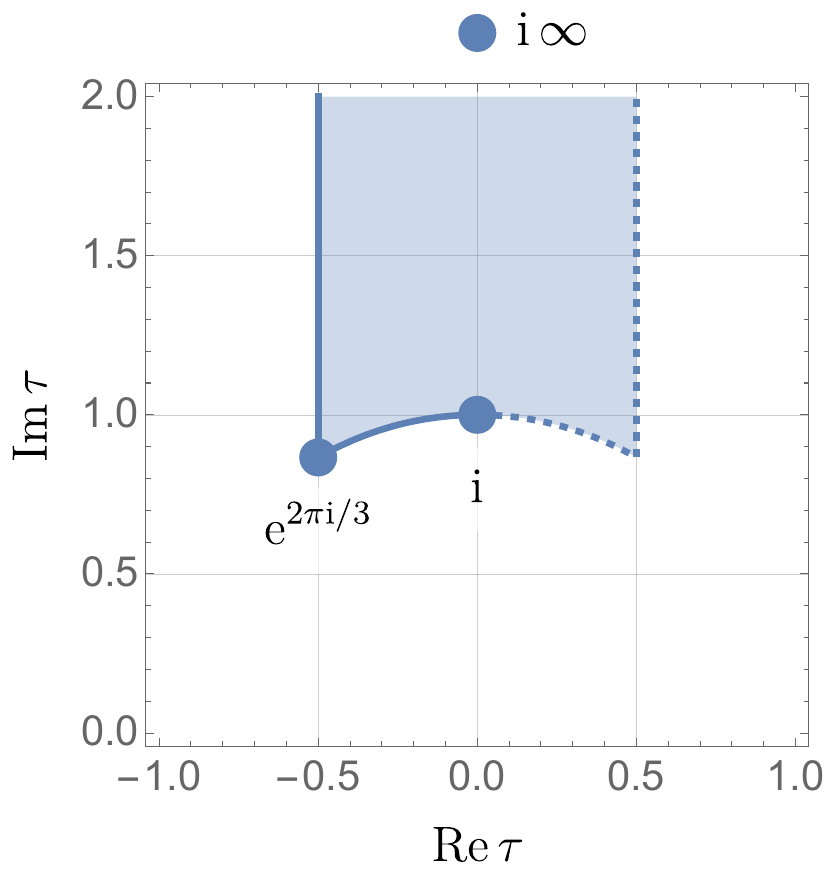}
    \caption{
    The fundamental domain of the $\SL(2, \mbb{Z})$ for the modulus $\tau$.
    The dots denote the three fixed points of the modular transformations.
    }
    \label{fig:fundamental-domain}
\end{figure}

By the actions of the generators $S$ and $T$, the modulus $\tau$ takes its value in the fundamental domain $\mc{F}$, which is defined by 
\begin{align}
    \mc{F} = \left\{
        \tau \in \mbb{C},
        \quad 
        \Im \tau >0,
        \quad 
        |\tau| > 1,
        \quad 
        |\Re \tau| < 1/2,
        \quad 
        \Re \tau = -1/2
    \right\},
\end{align}
where we decompose $\tau$ as $\tau = \Re \tau + \iu \Im \tau$.
This is the region shown in Fig.~\ref{fig:fundamental-domain}.
In this region, it is found that we have three fixed points for the modular transformation $\Gamma = \SL(2, \mbb{Z})$
\begin{align}
    \tau = \iu,
    \quad 
    \omega,
    \quad \iu \infty,
\end{align}
which are denoted by the dots in Fig.~\ref{fig:fundamental-domain}.
Here, $\omega \coloneqq \ee^{2 \pi \iu /3}$ being the cubic root of unity.
Once $\tau$ takes a value, the modular symmetry is spontaneously broken, but we obtain the following residual symmetries corresponding to the fixed points:
\begin{itemize}
    \item $\tau = \iu$: 
    This is the fixed point of the $S$ transformation and the residual symmetry is $\mbb{Z}^S_2$.
    \item $\tau = \omega$: 
    This is the fixed point of the $ST$ transformation and the residual symmetry is $\mbb{Z}^{ST}_3$.
    \item $\tau = \iu \infty$: 
    This is the fixed point of the $T$ transformation and the residual symmetry is $\mbb{Z}^T_N$.
\end{itemize}
These residual symmetries come from Eq.~\eqref{eq:STR-relations} and $T^N = \bm{1}$.

\subsubsection{Matter fields and modular forms}

Hereafter, we focus on the supergravity (SUGRA) formulation~\cite{Ferrara:1989bc,Wess:1992cp,Freedman:2012zz}.
Under the modular transformation $\gamma_N \in \Gamma_N$, a matter (super)field $X$ with the modular weight $k_X$ transforms as 
\begin{align}
    \gamma_N: X \mapsto (c \tau + d)^{k_X} \rho_X(\gamma) X,
    \label{eq:gamma-Phi}
\end{align}
where $\rho_X(\gamma)$ denotes the representation matrix determined by the representation of $X$ for $\Gamma_N$.
A modular form $Y(\tau)$ depending on $\tau$ similarly transforms under the modular transformation:
\begin{align}
	\gamma_N: Y \mapsto (c \tau+ d)^{k_Y} \rho_Y(\gamma) Y,
\end{align}
where $k_Y$ and $\rho_Y(\gamma)$ are the modular weight and representation matrix, respectively.
The modular weights $k_X$ and $k_Y$ are taken as even for simplicity.\footnote{
The weights can be taken also as odd, and then a modular group contains the non-trivial element of $R$.}
The matter K\"{a}hler potential is assumed to be given by 
\begin{align}
    \mc{K}_X = \frac{|X|^2}{(-\iu \tau + \iu \bar{\tau})^{-k_X}} = \frac{|X|^2}{(2 \Im \tau)^{-k_X}},
\end{align}
which is invariant under the transformations \eqref{eq:gamma-tau} and \eqref{eq:gamma-Phi}.

The K\"{a}hler potential for the modulus field $\tau$ typically has the following form
\begin{align}
    \mc{K}_\tau = - h \log (-\iu \tau + \iu \bar{\tau}).
\end{align}
Here, $h$ is a dimensionless constant, which is related to the choice of the compact space in the higher dimensional theory.
As shown in the following section, $h$ is irrelevant as long as we focus on the scalar potential generated by a coupling to the matter field $X$, and hence we continue our analysis without specifying the value of $h$.

In this paper, we focus on the case of $\Gamma_3 \simeq A_4$ $(N=3)$ and its modular forms for the concreteness of the model building.
For simplicity, we consider (non-)trivial singlet modular forms with the weight 4 and 6 in the follows. 
The discussion and expansion techniques presented below can similarly be applied to other modular forms.

\subsection{Expansion around fixed points}
\label{sec:Y-expansion}

Since we focus on the potential vacuum close to the fixed points via perturbation, 
we discuss techniques for expanding modular forms around each fixed point, in particular for $\tau = \iu$ and $\omega$.
The $q$-expansion of the modular form is useful for the fixed point $\tau = \iu \infty$, where $q = e^{2\pi \iu \tau}$.

\subsubsection{Expansion around $\tau = \iu$}
\label{sec:expansion-tau-iu}

$\tau = i$ is a fixed point associated with the $S$-transformation
\begin{align}
    S = \begin{pmatrix}
        0 & 1 \\
        -1 & 0
    \end{pmatrix},
    \qquad 
    S^2 = \bm{1},
\end{align}
and the residual symmetry is $\mbb{Z}_2^{S}$.
Under the $S$ transformation, the modular form $Y^{(k)}$ with the weight $k$ transforms as
\begin{align}
    Y^{(k)} \mapsto (-\tau)^k \rho_Y(S) Y^{(k)}
    \quad
    \text{for}
    \quad 
    \tau \mapsto - \frac{1}{\tau},
\end{align}
where $\rho_Y$ is a representation matrix of $Y^{(k)}$.
It is now convenient to introduce
\begin{align}
    s \coloneqq \frac{\tau - \iu}{\tau + \iu},
\end{align}
and to consider the modular forms as functions of $s$~\cite{Okada:2020ukr,Novichkov:2021evw,Feruglio:2022koo,Kikuchi:2023uqo}.
This parameter transforms as
\begin{align}
    S: s \mapsto -s,
\end{align}
under the $S$ transformation.
The modular form $Y^{(k)}$ here transforms as 
\begin{align}
    S: Y^{(k)}(s) \mapsto \biggl(
        \frac{1 + s}{1-s}
    \biggr)^k \iu^{-k}\rho_Y(S) Y^{(k)}(s).
\end{align}

Let us consider the singlets of $A_4$ as a concrete example.
The representation matrix becomes $\rho_Y(S) =1$, then 
\begin{align}
    Y^{(k)}(-s) = \biggl(
        \frac{1+s}{1-s}
    \biggr)^k \iu^{-k}Y^{(k)}(s).
    \label{eq:Stransformation-Y^k}
\end{align}
If we define 
\begin{align}
    \tilde{Y}^{(k)}(s) \coloneqq (1-s)^{-k} Y^{(k)}(s),
\end{align}
Eq.~\eqref{eq:Stransformation-Y^k} is rewritten in the following simple form:
\begin{align}
    \tilde{Y}^{(k)}(-s) = \iu^{-k} \tilde{Y}^{(k)}(s).
\end{align}
Considering the $l$-th derivative and taking $s \to 0$ of this equation, we obtain the useful relation~\cite{Kikuchi:2023uqo}
\begin{align}
    (\iu^{2l} - \iu^{-k}) \frac{\dd^l \tilde{Y}^{(k)}}{\dd s^l} \bigg|_{s \to 0} = 0.
    \label{eq:master-eq-tau-i}
\end{align}
We note that if the derivative of $Y^{(k)}$ has a non-vanishing value, the order of the derivatives $l$ and the modular weight $k$ need to satisfy $\iu^{2l} = \iu^{-k}$.
This tells us $Y^{(6)}_{\bm{1}} (\tau=i) = 0$, for instance.
With this relation and the chain rule
\begin{align}
    \frac{\dd s}{\dd \tau} = \frac{(1-s)^2}{2\iu},
\end{align}
we can derive the systematic expansion series of the singlet modular form around $\tau = \iu$.
For $k = 4$ and $6$, we can find 
\begin{align}
    Y^{(4)}_{\bm{r}}(\tau) &=
	\tilde{Y}^{(4)}_{\bm{r}}(0) 
	+ 2 \iu \tilde{Y}^{(4)}_{\bm{r}}(0) (\tau - \iu)
	+ \frac{1}{2} \biggl[
		- 5 \tilde{Y}^{(4)}_{\bm{r}} - \frac{1}{4} \frac{\dd^2 \tilde{Y}^{(4)}_{\bm{r}}}{\dd s^2}(0)
	\biggr] (\tau - \iu)^2
	\nn \\
	& \quad 
	- \frac{\iu}{3!} \biggl[
		15 \tilde{Y}^{(4)}_{\bm{r}}(0) + \frac{9}{4} \frac{\dd^2 \tilde{Y}^{(4)}_{\bm{r}}}{\dd s^2}(0)
	\biggr] (\tau - \iu)^3
	\nn \\
	& \quad 
	+ \frac{1}{4!} \biggl[
		\frac{105}{2} \tilde{Y}^{(4)}_{\bm{r}}(0) + \frac{63}{4} \frac{\dd^2 \tilde{Y}^{(4)}_{\bm{r}}}{\dd s^2}(0) + \frac{1}{16} \frac{\dd^4 \tilde{Y}^{(4)}_{\bm{r}}}{\dd s^4}(0)
	\biggr] (\tau - \iu)^4
	+ \mc{O}((\tau- \iu)^5) ,
    \label{eq:Y-expansion-i-4r}
    \\
    Y^{(6)}_{\bm{1}}(\tau) &=
	- \frac{\iu}{2} \frac{\dd \tilde{Y}^{(6)}_{\bm{1}}}{\dd s}(0) (\tau - \iu)
	+ \frac{7}{4} \frac{\dd \tilde{Y}^{(6)}_{\bm{r}}}{\dd s}(0) (\tau - \iu)^2
	\nn \\
	& \quad 
	+ \frac{\iu}{3!} \biggl[
		21 \frac{\dd \tilde{Y}^{(6)}_{\bm{1}}}{\dd s}(0) + \frac{1}{8} \frac{\dd^3 \tilde{Y}^{(6)}_{\bm{1}}}{\dd s^3}(0)
	\biggr] (\tau- \iu )^3
	\nn \\
	& \quad 
	+ \frac{1}{4!} \biggl[
		- 126 \frac{\dd \tilde{Y}^{(6)}_{\bm{1}}}{\dd s}(0) - \frac{9}{4} \frac{\dd^3 \tilde{Y}^{(6)}_{\bm{1}}}{\dd s^3}(0)
	\biggr] (\tau - \iu )^4 
	+ \mc{O}((\tau-\iu)^5),
    \label{eq:Y-expansion-i-61}
\end{align}
where $\bm{r} = \bm{1},~\bm{1}'$ for $k=4$.
Note that the modular forms in the left-handed side are functions of $\tau$ and that with a tilde in the right-handed side is a function of $s$.
See also App.~\ref{sec:result-taui-expansion} for the details.

\subsubsection{Expansion around $\tau = \omega$}
\label{sec:expansion-tau-omega}

$\tau = \omega$ is a fixed point associated with the $ST$ transformation
\begin{align}
    ST = \begin{pmatrix}
        0 & 1 \\
        -1 & -1
    \end{pmatrix},
    \qquad 
    (ST)^3 = \bm{1},
\end{align}
and the residual symmetry is $\mbb{Z}_3^{ST}$.
Under this transformation, a modular form $Y^{(k)}$ with the weight $k$ transforms as
\begin{align}
    Y^{(k)} \mapsto (-\tau-1)^k \rho_Y(ST) Y^{(k)}
\end{align}
and this becomes the following form if $\tau$ is set to be $\omega$:
\begin{align}
    Y^{(k)}(ST \omega) = \omega^{2k} \rho_Y(ST) Y^{(k)}(\omega).
    \label{eq:ST-transformation-f^k}
\end{align}

In a similar way to that in the expansion around $\tau = \iu$, it is convenient to introduce
\begin{align}
    u \coloneqq \frac{\tau - \omega}{\tau - \omega^2},
\end{align}
and consider the modular forms as the functions of $u$~\cite{Okada:2020ukr,Novichkov:2021evw,Petcov:2022fjf,Kikuchi:2023uqo}.
This parameter transforms as $u \mapsto \omega^2 u$.
Eq.~\eqref{eq:ST-transformation-f^k} is rewritten as
\begin{align}
    Y^{(k)}(\omega^2 u) = \biggl(
        \frac{1- \omega^2 u}{1- u}
    \biggr)^k \omega^{2k} \rho_Y(ST) Y^{(k)}(u).
    \label{eq:ST-transformation-Y}
\end{align}

If we introduce
\begin{align}
	\hat{Y}^{(k)} \coloneqq (1-u)^{-k} Y^{(k)},
\end{align}
Eq.~\eqref{eq:ST-transformation-Y} becomes
\begin{align}
    \hat{Y}^{(k)}(\omega^2 u) 
    = \omega^{2k} \rho_Y(ST) \hat{Y}^{(k)} = \omega^{-k} \rho_Y(ST) \hat{Y}^{(k)}(u).
\end{align}
From this equation, the $ST$ transformations for $k= 4, \ 6$ are given by 
\begin{align}
    \hat{Y}^{(4)}_{\bm{1}}(\omega^2 u) &= \omega^{-4}\hat{Y}^{(4)}_{\bm{1}}(u) = \omega^2 \hat{Y}^{(4)}_{\bm{1}}(u),
    \\
    \hat{Y}^{(4)}_{\bm{1}'} (\omega^2 u) &= \omega^{-3} \hat{Y}^{(4)}_{\bm{1}'}(u) = \hat{Y}^{(4)}_{\bm{1}'}(u),
    \\
    \hat{Y}^{(6)}_{\bm{1}} (\omega^2 u) &= \omega^{-6} \hat{Y}^{(6)}_{\bm{1}}(u) = \hat{Y}^{(6)}_{\bm{1}}(u),
\end{align}
where we use $\rho_{\bm{1}}(ST) =1$ and $\rho_{\bm{1}'}(ST) = \omega$.
Considering the $l$-th derivatives of these relations and taking $u \to 0$, we obtain the following useful relation like Eq.~\eqref{eq:master-eq-tau-i}
\begin{align}
    (\omega^{2l} - \omega^{q_{\bm{r}} - k}) \frac{\dd^l \hat{Y}^{(k)}_{\bm{r}}}{\dd u^l} \bigg|_{u \to 0} = 0,
    \qquad
    q_{\bm{r}} = \begin{cases}
        0 & \bm{r} = \bm{1}
        \\
        1 & \bm{r} = \bm{1}'
    \end{cases}.
\end{align}
It is noted that if $2l \equiv q_{\bm{r}} - k \mod 3$, the derivative of the modular form can take a non-zero value.
Using the chain rule 
\begin{align}
    \frac{\dd u}{\dd \tau} = \frac{(1-u)^2}{\sqrt{3} \iu},
\end{align}
we can get the following systematic expansions:
\begin{align}
    Y^{(4)}_{\bm{1}}(\tau) &= 
	- \frac{\iu}{\sqrt{3}} \frac{\dd \hat{Y}^{(4)}_{\bm{1}}}{\dd u}(0) (\tau - \omega)
	+ \frac{5}{3} \frac{\dd \hat{Y}^{(4)}_{\bm{1}}}{\dd u}(0) (\tau - \omega)^2
	\nn \\
	& \quad 
	+\frac{\iu}{3!} \biggl[
		\frac{30}{\sqrt{3}} \frac{\dd \hat{Y}^{(4)}_{\bm{1}}}{\dd u}(0)
	\biggr](\tau - \omega)^3
	\nn \\
	& \quad 
	+ \frac{1}{4!} \biggl[
		- \frac{280}{3} \frac{\dd \hat{Y}^{(4)}_{\bm{1}}}{\dd u}(0) + \frac{1}{9} \frac{\dd^4 \hat{Y}^{(4)}_{\bm{1}}}{\dd u^4}(0)
	\biggr] (\tau - \omega)^4 
	+ \mc{O}((\tau- \omega)^5),
    \label{eq:Y-expansion-omega-41}
    \\
    Y^{(4)}_{\bm{1}'}(\tau) &= 
	\hat{Y}^{(4)}_{\bm{1}'}(0)
	+ \frac{4 \iu}{\sqrt{3}} \hat{Y}^{(4)}_{\bm{1}'}(0)  (\tau - \omega)
	- \frac{10}{3} \hat{Y}^{(4)}_{\bm{1}'}(0)(\tau - \omega)^2
	\nn \\
	& \quad 
	+\frac{\iu}{3!} \biggl[
		- \frac{40}{\sqrt{3}} \hat{Y}^{(4)}_{\bm{1}'}(0) + \frac{1}{3 \sqrt{3}} \frac{\dd^3 \hat{Y}^{(4)}_{\bm{1}'}}{\dd u^3} (0)
	\biggr] (\tau - \omega)^3
	\nn \\
	& \quad 
	+ \frac{1}{4!} \biggl[
		\frac{280}{3} \hat{Y}^{(4)}_{\bm{1}'}(0) - \frac{28}{9} \frac{\dd^3 \hat{Y}^{(4)}_{\bm{1}'}}{\dd u^3}(0)
	\biggr] (\tau - \omega)^4
	+ \mc{O}((\tau - \omega)^5),
    \label{eq:Y-expansion-omega-41p}
    \\
    Y^{(6)}_{\bm{1}}(\tau) & = 
	\hat{Y}^{(6)}_{\bm{1}}(0)
	+ 2\sqrt{3} \iu \hat{Y}^{(6)}_{\bm{1}}(0) (\tau - \omega)
	- 7 \hat{Y}^{(6)}_{\bm{1}}(\tau - \omega)^2
	\nn \\
	& \quad 
	+ \frac{\iu}{3!} \biggl[
		- \frac{112}{\sqrt{3}} \hat{Y}^{(6)}_{\bm{1}}(0) + \frac{1}{3 \sqrt{3}} \frac{\dd^3 \hat{Y}^{(6)}_{\bm{1}}}{\dd u^3}(0)
	\biggr] (\tau - \omega)^3
	\nn \\
	& \quad 
	+ \frac{1}{4!}\biggl[
		336 \hat{Y}^{(6)}_{\bm{1}}(0) - 4 \frac{\dd^3 \hat{Y}^{(6)}_{\bm{1}}}{\dd u^3}(0)
	\biggr] (\tau - \omega)^4
	+ \mc{O}((\tau-\omega)^5).
    \label{eq:Y-expansion-omega-61}
\end{align}
The modular form in the left-handed side is a function of $\tau$ and that with a hat in the right-handed side is a function of $u$.
Note that $Y^{(4)}_{\bm{1}} (\tau=\omega) = 0$, for instance.
See also App.~\ref{sec:result-tauomega-expansion} for the details of the derivation.

\subsubsection{Expansion around $\tau = \iu \infty$}

At $\tau = \iu \infty$, the so-called $q$-expansion works well, where $q \coloneqq \exp(2\pi \iu \tau)$~\cite{Okada:2020ukr,Feruglio:2022koo,Kikuchi:2023uqo,Novichkov:2021evw}.
The leading order expansions of the trivial and non-trivial singlets with the modular weights 4 and 6 are given by 
\begin{align}
    Y^{(4)}_{\bm{1}} &\approx 1 + 240 q,
    \\
    Y^{(4)}_{\bm{1}'} & \approx -12 q^{1/3},
    \label{eq:q-expansoin-Y41p}
    \\
    Y^{(6)}_{\bm{1}} &\approx 1 - 504 q.
\end{align}
Note that $Y^{(4)}_{\bm{1}'}(\tau = \iu \infty)= 0$ 
since $q \to 0$ in the large $\Im \tau$ limit.
See also App.~\ref{sec:result-q-expansion} for the higher order terms and their derivatives.

\section{Model in supergravity}
\label{sec:Model}

\subsection{Scalar potential}

In this section, we introduce the model and its scalar potential.
We focus on the moduli potential of the F-term SUGRA potential generated by a coupling to the matter field $X$.
The F-term potential in SUGRA is given by~\cite{Wess:1992cp,Freedman:2012zz}
\begin{align}
    V = \ee^{\mc{K}} \Bigl[
        \mc{K}^{i \bar{j}} D_i W \ol{D_j W} - 3 |W|^2
    \Bigr]
    \label{eq:sugra-potential-formula}
\end{align}
where $\mc{K}_{i\bar{j}}$ is the total K\"{a}hler metric and 
$\mc{K}^{i \bar{j}}$ denotes its inverse.
$D_i$ acts on the superpotential as $D_i W = \del_i W + (\del_i \mc{K}) W$ with the K\"{a}hler potential $\mc{K}$.
The indices $i, j$ run all the superfield components.
The total K\"{a}hler potential is assumed to be
\begin{align}
    \mc{K} = - h \log (-\iu \tau + \iu \bar{\tau}) + \sum_a^{N_a} \biggl[
    		\frac{|X_a|^2}{(- \iu \tau + \iu \bar{\tau})^{-k_{X_a}}}
    		- \frac{|X_a|^4}{4 \Lambda_{X_a}^2 (- \iu \tau + \iu \bar{\tau})^{-2 k_{X_a}}}
    	\biggr]
    	\label{eq:total-Kahler-potential}
\end{align}
where the index $a$ runs all the matter fields $X_a$, $k_{X_a}$ is the modular weight of $X_a$, $N_a$ is the total number of $X_a$, and we choose $h=2$ in the numerical analysis of this paper.
$|X_a|^4$ terms are introduced to stabilizer $X_a$ and their $\tau$-dependent coefficients are chosen such that the action is modular invariant.
$\Lambda_{X_a}$ is an energy scale that suppresses the higher order terms.
We consider the following superpotential
\begin{align}
	W = \Lambda^2 \sum_a^{N_a} c_a Y_a^{(k_{Y_a})} X_a,
	\label{eq:total-superpotential}
\end{align}
where $Y_a^{(k_{Y_a})}$ is a modular form with the weight $k_{Y_a}$ and $c_a$ is a numerical coefficient.
$\Lambda$ is a typical energy scale of the superpotential and we set $\Lambda= 1$ for simplicity hereafter.
These $X_a$'s are so-called stabilizer fields~\cite{Kallosh:2010ug,Kallosh:2010xz,Kallosh:2014vja,Ferrara:2014kva,Higaki:2016ydn}, which are matter fields introduced to generate a scalar potential for a modulus field~\cite{Knapp-Perez:2023nty}.
Here, we assume that the representation of $X_a$ is a singlet
such that the superpotential is modular invariant.
This system has an $R$-symmetry under the phase transformation:
\begin{align}
	\rU(1)_R: X_a \mapsto \exp(2 \iu \alpha) X_a,
\end{align}
where $\alpha$ is a constant transformation parameter.
This symmetry controls the scalar potential as shown in the following sections.
The modular weights $k_{X_a}$ and $h$ satisfy the following relation 
\begin{align}
	-h = k_{Y_a} + k_{X_a},
	\label{eq:modular-weight-relation}
\end{align}
for each label $a$ \cite{Ohki:2020bpo,Abe:2023ylh}.
If $X_a$ are much smaller than the Planck scale, i.e. $|X_a|^2 \ll 1$, the scalar potential \eqref{eq:sugra-potential-formula} is evaluated as 
\begin{align}
	V = \sum_{a}^{N_a} (-\iu \tau + \iu \bar{\tau})^{k_{Y_a}} |c_a Y^{(k_{Y_a})}|^2 +{\cal O}(|X|^2),
\end{align}
where we use \eqref{eq:modular-weight-relation}. 
Hereafter, we study the moduli stabilization including $X_a$ with $N_a = 1$ and $N_a = 2$ in Sec.~\ref{sec:single-modular-form-potential} and Sec.~\ref{sec:double-modular-forms-potential}, respectively.

\subsection{Moduli stabilization and overview of our models}
\label{sec:moduli-stabilization}

Before focusing on the analysis of the $\tau$ vacuum, let us consider the properties of the stabilizer fields in Eqs.~\eqref{eq:total-Kahler-potential} and \eqref{eq:total-superpotential}.
We give also overviews of our models.

\subsubsection{Single stabilizer case}

First, we start the discussion with a single stabilizer case.
The K\"{a}hler potential and superpotential are given by
\begin{align}
    \mc{K} &= Z_X|X|^2 
    - \frac{Z_X^2|X|^4}{4 \Lambda_X^2}
    - h \log ( - \iu \tau + \iu \bar{\tau}),
    \qquad 
    Z_X = \frac{1}{(- \iu \tau + \iu \bar{\tau})^{-k_X}},
    \\
    W &= Y(\tau) X.
\end{align}
If we focus on the $X$-dependence in the scalar potential, it is expanded as 
\begin{align}
    V &=(2\Im \tau )^{-(h+k_X)} |Y(\tau)|^2 
    \nn \\
    & \quad 
    + (2 \Im \tau)^{-h} \biggl[
        \frac{|Y|^2}{\Lambda^2_X} + \frac{|(h+k_X) Y(\tau) - 2 \iu \Im \tau \del_\tau Y(\tau)|^2 }{h}
    \biggr] |X|^2 + \mc{O}(|X|^4)
    \\
    &\eqqcolon V_0 + m_X^2 |X|^2 + \mc{O}(|X|^4).
\end{align}
$m_X^2$ is positive definite if the modulus is stabilized such that conditions $Y \neq 0$ or $\del_\tau Y \neq 0$ are satisfied.
Then we obtain $\braket{X} = 0$ and this is the case in this paper, except for $Y=Y^{(4)}_{{\bm 1}'}$.

The potential $V_0 \coloneqq (2 \Im \tau)^{-(h+k_X)} |Y|^2$ gives a de Sitter (dS) vacuum or a supersymmetric one depending on $\braket{Y(\tau)}$:
\begin{align}
    \begin{cases}
        \displaystyle
        \braket{V_0} > 0 \ \text{for} \ \braket{Y} \neq 0 \quad \Rightarrow \quad \text{dS vacuum}
        \\
        \\
        \displaystyle
        \braket{V_0} = 0 \ \text{for} \ \braket{Y} = 0 \quad \Rightarrow \quad \text{supersymmetric Minkowski vacuum}
    \end{cases}.
\end{align}
The modulus $\tau$ develops the VEV due to the zero point value of $Y$ because of $V_0 \propto |Y|^2$.
Then $Y =0$ and $\del_\tau Y \neq 0$ are realized in the vacuum.
We note that $\braket{W} = 0$ is always satisfied for $\braket{X} = 0$ and the F-term is given by $F_X \sim \del_X W \sim Y =0$. 
Then we obtain the supersymmetric Minkowski vacuum with $\braket{Y} = 0$ in a single $X$ model in this paper.

We note that the canonically normalized mass squared is given by $m_X^2 / \mc{K}_{X\bar{X}} = m_X^2 / Z_X$.
$\mc{K}_{X\bar{X}}$ is the coefficient of the kinetic term of $X$ and the explicit form of the K\"{a}hler metric is shown in App.~\ref{sec:Kahler-metric}.

\paragraph{Comment on the $R$-symmetry violation.}

Let us consider adding a $\tau$-dependent term $W_0(\tau)$ to the superpotenial as 
\begin{align}
	W = Y(\tau) X + W_0(\tau),
\end{align}
where $W_0(\tau)$ depends only the modulus $\tau$ \cite{Lebedev:2006qq,Dudas:2006gr,Abe:2006xp,Kallosh:2006dv,Abe:2007yb,Knapp-Perez:2023nty}.\footnote{See also Ref.\cite{Kachru:2003aw}.}
This $W_0$ will generally exist in the quantum gravity since it violates the global $R$-symmetry, and it induces an additional contribution to the scalar potential.
Then, the potential is roughly given by 
\begin{align}
	V \sim \bigl[ V_0 - 3 |W_0|^2 \bigr] + m_X^2  |X|^2 - 3 W_0 \bar{Y} \bar{X} - 3 \ol{W_0} Y X + \cdots .
\end{align}
Here, a simple estimate of vacua in several models will be given below. 
As noted, a choice of $Y$ controls properties of $V$. 
For instance, there can be models which involves either $Y$ such that $Y=0$ at a value of $\tau$ or $Y\neq 0$ for any values of $\tau$.
In addition, there may be cases in which $\langle Y\rangle \neq 0$ in the vacuum due to the correction term of $W_0$. 
Thus, $\braket{X}$ becomes as follows depending on a choice of $Y$:
\begin{align}
	\braket{X} \sim \begin{cases}
		\displaystyle
		\frac{\Lambda_X^2}{\braket{Y}} W_0 & \text{for } \braket{Y} \neq 0,
		\\
		\\
		\displaystyle
		0 & \text{for } \braket{Y} = 0.
	\end{cases}
\end{align}
where $m_X^2 \sim |Y|^2/\Lambda_X^2$ is used for $\braket{Y} \neq 0$.
Note that $\partial_\tau Y$ contributes to $m_X^2$.
If $\braket{Y} =0$ or $|\braket{Y}| \gg |W_0|$, the VEV of $X$ remains tiny as discussed above. 
A shift of $\tau$ VEV will be also small and, then a smaller cosmological constant $\braket{V} \sim \braket{V_0 - 3 |W_0|^2}$ is obtained.

On the other hand, for the case of $|Y| \sim |W_0|$, it will be necessary to study the detail of a modulus stabilization including the $W_0$ because we roughly evaluate
\begin{align}
	D_X W \sim Y \sim W_0,
	\quad 
	D_\tau W \sim D_\tau W_0 \sim W_0,
	\qquad 
	W = Y X + W_0 \sim W_0.
\end{align}
Then the magnitude of the F-term scalar potential is characterized by $W_0$ \cite{Lebedev:2006qq,Dudas:2006gr,Abe:2006xp,Kallosh:2006dv,Abe:2007yb,Knapp-Perez:2023nty}.
However, a concrete study is beyond our scope in this paper.

\subsubsection{Double stabilizer case}
\label{sec:DS-case}

Next, let us discuss the model with two matter fields, say $X$ and $\Xi$.
In this model, the K\"{a}hler potential and superpotential are given by
\begin{align}
	\mc{K} &= Z_X |X|^2 - \frac{Z_X^2|X|^4}{4 \Lambda_X^2} + Z_\Xi |\Xi|^2 - \frac{Z_\Xi |\Xi|^4}{4 \Lambda_\Xi^2} - h \log ( - \iu \tau + \iu \bar{\tau}),
    \\
    Z_X & = \frac{1}{( - \iu \tau + \iu \bar{\tau})^{-k_X}},
    \qquad 
    Z_\Xi = \frac{1}{( - \iu \tau + \iu \bar{\tau})^{-k_\Xi}},
    \\
    W & = Y_X(\tau) X + Y_\Xi(\tau) \Xi,
\end{align}
where an additional stabilizer field is denoted by $\Xi$.
The mixing between $X$ and $\Xi$ vanishes due to the (approximate) $\rU(1)_R$ symmetry.
By Eq.~\eqref{eq:sugra-potential-formula}, we obtain
\begin{align}
    V &= (2 \Im \tau)^{-(h+k_X)} |Y_X(\tau)|^2 + (2 \Im \tau)^{-(h+k_\Xi)} |Y_\Xi(\tau)|^2
    \nn \\
    &\quad
    +(2  \Im \tau)^{-h} \biggl[
        \frac{|Y_X(\tau)|^2}{\Lambda_X^2}
        +|Y_\Xi(\tau)|^2 (2  \Im \tau)^{-k_\Xi + k_X}
        +\frac{|(h + k_X) Y_X(\tau) - 2 \iu \Im \tau \del_\tau Y_X(\tau)|^2}{h}
    \biggr] |X|^2
    \nn \\
    & \quad 
    + (2  \Im \tau)^{-h} \biggl[
        \frac{|Y_\Xi(\tau)|^2}{\Lambda_\Xi^2} 
        + |Y_X(\tau)|^2 (2 \Im\tau)^{-k_X + k_\Xi} 
        + \frac{|(h+ k_\Xi)Y_\Xi(\tau) - 2 \iu \Im \tau \del_\tau Y_\Xi(\tau)|^2}{h}
    \biggr] |\Xi|^2
    \nn \\
    & \quad 
    + \frac{(2  \Im\tau)^{-h}}{h} \biggl[
        \Bigl\{
            \bigl( (h+ k_X) Y_X(\tau) - 2 \iu \Im \tau \del_\tau Y_X \bigr) \bigl( (h+ k_\Xi) \ol{Y_\Xi(\tau)} + 2 \iu \Im \tau \ol{\del_\tau Y_\Xi(\tau)} \bigr) 
            \nn \\
            & \qquad - h Y_X(\tau) \ol{Y_\Xi(\tau)}
        \Bigr\} X \bar{\Xi} + \mr{c.c.}
    \biggr]
    + \cdots
    \label{eq:DMF-potential-formula}    \\
    &\eqqcolon V_0 + m_X^2 |X|^2 + m_\Xi^2 |\Xi|^2 + (M^2 X \bar{\Xi} + \mr{c.c.}) +\cdots,
\end{align}
where we expand the scalar potential to the second order of stabilizers.
The ellipsis denotes the higher order terms in $X$ and $\Xi$.
The quadratic term in this potential is expressed as 
\begin{align}
    V \ni \begin{pmatrix}
        \bar{X} & \bar{\Xi}
    \end{pmatrix}
    \begin{pmatrix}
        m_X^2 & \ol{M^2} \\
        M^2 & m_\Xi^2
    \end{pmatrix} \begin{pmatrix}
        X \\
        \Xi
    \end{pmatrix}.
    \label{eq:mass-matrix}
\end{align}
We note that this simple form results from the $\rU(1)_R$ symmetry.
Then we obtain the following two mass eigenvalues from this mass squared matrix:
\begin{align}
	m_\pm^2 = \frac{1}{2}\left[
		\frac{m_X^2}{\mc{K}_{X\bar{X}}}
		+ \frac{m_\Xi^2}{\mc{K}_{\Xi\bar{\Xi}}}
		\pm \sqrt{
			4 \frac{|M|^4}{\mc{K}_{X\bar{X}} \mc{K}_{\Xi\bar{\Xi}}}
			+ \left( \frac{m_X^2}{\mc{K}_{X\bar{X}}} - \frac{m_\Xi^2}{\mc{K}_{\Xi \bar{\Xi}}} \right)^2
		}
	\right].
\end{align}
Here we normalize the mass term of Eq.~\eqref{eq:mass-matrix} with the K\"{a}hler metric so that the stabilizer fields have canonically normalized kinetic terms.
For details of the K\"{a}hler metric, see App.~\ref{sec:Kahler-metric}.
If the stabilizer fields are stabilized at $\braket{X} = \braket{\Xi} =0$, the K\"{a}hler metric reduces to $\mc{K}_{X\bar{X}} = Z_X$ and $\mc{K}_{\Xi\bar{\Xi}} = Z_\Xi$, and the mass eigenvalues become
\begin{align}
	m_\pm^2 = \frac{1}{2} \left[
		\frac{m_X^2}{Z_X} + \frac{m_\Xi^2}{Z_\Xi} \pm \sqrt{
			4 \frac{|M|^4}{Z_X Z_\Xi} + \left( \frac{m_X^2}{Z_X} - \frac{m_\Xi^2}{Z_\Xi} \right)^2
		}
	\right].
 \label{eq:X-mass-eigen}
\end{align}

Depending on the VEVs of the modular forms, the following three ways can be considered in general:
\begin{enumerate}
    \item $\braket{Y_X} \neq 0, ~\braket{Y_\Xi} \neq 0$.\\
    The mass squared parameters in Eq.~\eqref{eq:mass-matrix} satisfy
    \begin{align}
        m_X^2 > m_\Xi^2 \gtrsim |M|^2,
    \end{align}
    for $\Lambda_X \sim \Lambda_\Xi \lesssim 1$, and the mass eigenvalues become approximately
    \begin{align}
        m_+^2 \approx \frac{m_X^2}{\mc{K}_{X\bar{X}}} + \frac{|M|^4/\mc{K}_{X\bar{X}} \mc{K}_{\Xi\bar{\Xi}}}{m_X^2/\mc{K}_{X\bar{X}} - m_\Xi^2 / \mc{K}_{\Xi\bar{\Xi}}},
	\qquad 
	m_-^2 \approx \frac{m_\Xi^2}{\mc{K}_{\Xi\bar{\Xi}}} - \frac{|M|^4/\mc{K}_{X\bar{X}} \mc{K}_{\Xi\bar{\Xi}}}{m_X^2/\mc{K}_{X\bar{X}} - m_\Xi^2 /\mc{K}_{\Xi\bar{\Xi}}},
    \end{align}
    which leads to $m_\pm^2 \geq 0$.
    From this, the vacuum is characterized by $\braket{X} = \braket{\Xi} = 0$.
    The vacuum is dS because of $V \sim |Y_X|^2 + |Y_\Xi|^2 \neq 0$.
    \item $\braket{Y_X} \neq 0,~\braket{Y_\Xi} = 0$.\\
    The mass squared parameters have the following relation:
    \begin{align}
        m_X^2 \gtrsim m_\Xi^2 \sim |M|^2,
    \end{align}
    for $\Lambda_X \sim \Lambda_\Xi \lesssim 1$.
    Even in this case, the lighter mass squared eigenvalue is positive definite;
    \begin{align}
        m_-^2 \approx \frac{m_\Xi^2}{\mc{K}_{\Xi\bar{\Xi}}} - \frac{|M|^4}{m_X^2 \mc{K}_{\Xi\bar{\Xi}}} \approx \frac{m_\Xi^2}{\mc{K}_{\Xi\bar{\Xi}}} = \frac{m_\Xi^2}{Z_\Xi} \geq 0,
    \end{align}
    where we assume $\braket{X} = \braket{\Xi} =0$ here.
    The vacuum is dS because of $V \sim |Y_X|^2 \neq 0$.
    \item $\braket{Y_X} = \braket{Y_\Xi} =0$.\\
    If the derivatives of the modular forms do not vanish, we obtain non-vanishing mass squared terms for the stabilizer fields and a vanishing cosmological constant $V = 0$.
\end{enumerate}
The cases 1 and 2 are realized in the following concrete models and 3 is not the case in this paper.
dS or Minkowski vacua are obtained if we have a multi-modular form contribution.

Similarly to the single $X$ case, $W_0(\tau)$ which violates $R$-symmetry explicitly can be added to the superpotential.
It is expected that the vacuum energy can be changed as discussed already, but a concrete study is beyond our scope in this paper.

\section{Single modular form potential}
\label{sec:single-modular-form-potential}

\begin{figure}
    \centering
    \begin{subfigure}{0.31\textwidth}
        \centering
        \includegraphics[width=\textwidth]{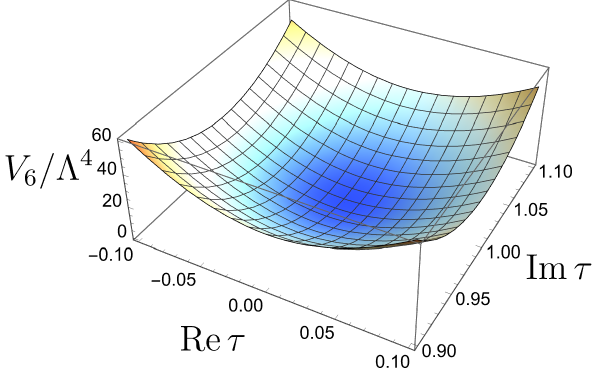}
        \caption{$V^{(6)}_{\bm{1}}$}
    \end{subfigure}
    \begin{subfigure}{0.31\textwidth}
        \centering
        \includegraphics[width=\textwidth]{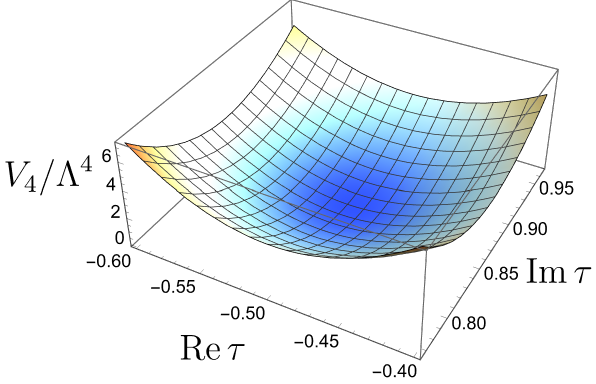}
        \caption{$V^{(4)}_{\bm{1}}$}
    \end{subfigure}
    \begin{subfigure}{0.31\textwidth}
        \centering
        \includegraphics[width=\textwidth]{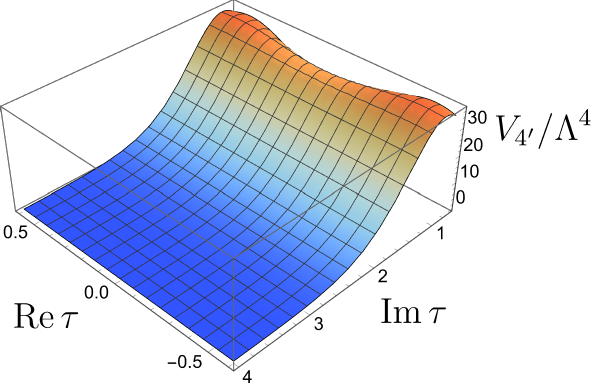}
        \caption{$V^{(4)}_{\bm{1}'}$}
    \end{subfigure}
    \caption{
    3D plots of the single modular form potential \eqref{eq:single-form-potential}.
    These potentials have their global vacua at $\tau = \iu,~\omega,$ and $\iu \infty$ on the fundamental domain, respectively. The points are fixed points of the modular transformation.
    }
    \label{fig:f-VY}
\end{figure}

First, let us start with the vacuum analysis of the single modular form potential.
In this case, the superpotential has a simple form 
\begin{align}
    W^{(k_Y)}_{\bm{r}} = Y^{(k_Y)}_{\bm{r}} X_{-h - k_Y},	
\end{align}
and K\"{a}hler metric is given by Eq.~\eqref{eq:Kahler-metric-single-modular-form}.
$k_Y$ and $\bm{r}$ denote the modular weight and the representation of the singlet modular form $Y^{(k_Y)}_{\bm{r}}$, respectively.
As examples, we will consider cases with using
\begin{align}
    Y^{(k_Y)}_{\bm{r}} = Y^{(4)}_{\bm{1}},~Y^{(4)}_{\bm{1}'},
    ~\text{or} ~
    Y^{(6)}_{\bm{1}}.
\end{align}
Here, $Y^{(k_Y)}_{\bm{r}}=0$ at $\tau= \omega,~\iu \infty,$ or $\iu$, respectively, as discussed in Sec.~\ref{sec:Y-expansion}, which are also fixed points of the modular transformation.
$X$ has the modular weight $-h - k_Y$ from Eq.~\eqref{eq:modular-weight-relation}.
A representation of $X$ is a trivial singlet when $X$ is coupled to $Y^{(4)}_{\bm{1}}$ or $Y^{(6)}_{\bm{1}}$, while it is a non-trivial singlet $\bm{1}''$ when $X$ is coupled to $Y^{(4)}_{\bm{1}'}$.
The scalar potential is given by 
\begin{align}
    V^{(k_Y)}_{\bm{r}} \coloneqq (2 \Im \tau)^{k_Y} |Y^{(k_Y)}_{\bm{r}}|^2.
    \label{eq:single-form-potential}
\end{align}
Thus, the minimum of the scalar potential exists at $\tau= \omega,~\iu \infty,$ or $\iu$ in the fundamental domain when we take $Y^{(k_Y)}_{\bm{r}} = Y^{(4)}_{\bm{1}},~Y^{(4)}_{\bm{1}'} {\rm or}~Y^{(6)}_{\bm{1}}$, respectively.

$\tau$ and $X$ are stabilized in the models with the potential constructed by $Y^{(4)}_{\bm{1}}$ or $Y^{(6)}_{\bm{1}}$.
Recovering the overall scale of $\Lambda$ and the Planck scale, we obtain the modulus mass squared $m_\tau^2 \sim \partial_\tau \partial_{\bar{\tau}} V/\mc{K}_{\tau\bar{\tau}}$, where $\mc{K}_{\tau\bar{\tau}}= h M_{P}^2/(2{\rm Im}\tau)^2$, at the respective supersymmetric Minkowski vacuum\footnote{
The $M_P$-dependence in the modulus mass will not be shown explicitly in the following sections.}:
\begin{align}
    m_\tau \sim \frac{\Lambda^2}{M_P} |\partial_\tau Y^{(k)}_{\bm 1}(\langle \tau\rangle)| \sim 10
    \frac{\Lambda^2}{M_P}.
\end{align}
Here, $\langle \tau \rangle = \omega$ and $\iu$ for $k=4$ and $6$, respectively.
On the other hand, the potential constructed by $Y^{(4)}_{\bm{1}'}$ is a kind of runaway potential because $V^{(4)}_{\bm{1}'} \sim |q|^{2/3}$ in the $q$-expansion as shown in Eq.~\eqref{eq:q-expansoin-Y41p}. 
Then, there exists no vacuum at a finite value of $\tau$ on the fundamental domain and therefore $X$ is not stabilized owing to the vanishing mass squared at $\tau = \iu \infty$.

The 3D plots of the scalar potentials $V^{(6)}_{\bm{1}}$, $V^{(4)}_{\bm{1}}$, and $V^{(4)}_{\bm{1}'}$ are shown in Fig.~\ref{fig:f-VY}.
The potentials (a) and (b) are found to have the minimum at $\tau = \iu$ and $\tau = \omega$, respectively.
The potential (c) with $Y^{(4)}_{\bm{1}'}$ is runaway as mentioned above.

The modulus VEVs at the fixed points are useful for particle models, since the models with a finite modular symmetry for a flavor hierarchy often require the VEVs nearby the fixed points \cite{Okada:2020ukr,Feruglio:2021dte,Novichkov:2021evw,Petcov:2022fjf,Kikuchi:2023cap,Abe:2023ilq,Petcov:2023vws,Abe:2023qmr,Abe:2023dvr,deMedeirosVarzielas:2023crv,Kikuchi:2023dow,Kikuchi:2023fpl}.
On the other hand, a small deviation of the VEV from the fixed point is necessary for a realistic flavor hierarchy, and hence we will consider the deformation of the potential by adding another modular form contribution.

\section{Double modular forms potential}
\label{sec:double-modular-forms-potential}

We discuss modulus stabilization with two modular forms in the potential.
The superpotential in each model is given by
\begin{align}
	W_{6+4} &\coloneqq Y^{(6)}_{\bm{1}} X_{-h-6} + c Y^{(4)}_{\bm{1}} X_{-h-4},
	\nn \\ 
	W_{6 +4'} &\coloneqq Y^{(6)}_{\bm{1}} X_{-h-6} + c Y^{(4)}_{\bm{1}'} X_{-h-4}',
	\\
	W_{4 +4'} &\coloneqq Y^{(4)}_{\bm{1}} X_{-h-4} + c Y^{(4)}_{\bm{1}'} X_{-h-4}',
	\nn
\end{align}
where $c$ is a dimensionless parameter controlling an additional contribution of a modular form.
The index of $X$ denotes the modular weight of the stabilizer field.
$X_{-h-4}$ and $X_{-h-6}$ are trivial singlets, whereas $X_{-h-4}'$ is $\bm{1}''$.
Here, suppose that $X$'s are stabilized at the origin, and we will confirm it numerically in Sec.~\ref{sec:X-mass}.
Then, the scalar potential is given in the following form as discussed in Eq.~(\ref{eq:DMF-potential-formula}):
\begin{align}
    V_{6+4} \coloneqq V^{(6)}_{\bm{1}} + c^2 V^{(4)}_{\bm{1}},
    \quad 
    V_{6 + 4'} \coloneqq V^{(6)}_{\bm{1}} + c^2 V^{(4)}_{\bm{1}'},
    \quad
    V_{4 + 4'} \coloneqq V^{(4)}_{\bm{1}} + c^2 V^{(4)}_{\bm{1}'},
    \label{eq:scalar-potential-V_r}
\end{align}
where $V^{(k_Y)}_{\bm{r}}$ is a single modular form potential defined by Eq.~\eqref{eq:single-form-potential}.
We also use the notation $V_{r}\ (r = 6+4,\ 6+4', \ 4+ 4')$.

In the rest part of this section, we show 3D plots of these scalar potentials and their vacuum shifts from a fixed point. 
To this end, we consider a perturbation of a $c$-dependent term to a fixed point in order to exhibit instabilities via results developed in Sec.~\ref{sec:Y-expansion}.

\subsection{Brief summary}

\begin{table}[t]
	\centering
	\begin{subtable}[c]{0.5\textwidth}
		\centering
		\begin{tabular}{c|c|c|c}\hline
  & $c\lesssim 2$ & $ c \gtrsim 2$ & \\
			\hline
			Vacuum & $\braket{\tau} = \iu$ & $\braket{\tau} = \omega$ & dS \\
			\hline
		\end{tabular}
		\caption{$V_{6+4}$}
	\end{subtable}

	\begin{subtable}[c]{0.5\textwidth}
		\centering
		\begin{tabular}{c|c|c|c|c}\hline
  		& $c \lesssim 5$ & $ c\gtrsim 5$ & large $c$ & \\
			\hline
			Vacuum & $\braket{\tau} = \iu$ & $ \braket{\tau} = \iu ( 1 + y(c))$& $ \braket{\tau} = \iu y(c)$ & dS\\
			\hline
		\end{tabular}
		\caption{$V_{6+4'}$}
	\end{subtable}

	\begin{subtable}[c]{0.5\textwidth}
		\centering
		\begin{tabular}{c|c|c|c}\hline
	& $c \lesssim 2$ & $c\gtrsim 2$ & \\		
			\hline
			Vacuum & $\braket{\tau} = \omega$ &
   $\braket{\tau} = -1/2+ \iu y(c)$
   & dS \\
			\hline
		\end{tabular}
		\caption{$V_{4+4'}$}
	\end{subtable}
	\caption{
	A brief summary of the vacua and $c$ values corresponding to each vacuum.
    $\braket{\tau} = \iu$ and $\omega$ are the fixed points and give CP-conserving vacua.
    Real numbers of $y(c)$'s in the table (b) and (c) are fluctuations of $\Im(\tau)$ around fixed points. 
    $y(c)$ has non-trivial $c$ dependence.
    When $c\gtrsim 5$ in (b), $y(c)<1$ is given by the stationary condition of Eq.~\eqref{eq:yofc-V6+4p-taui-2}.
    For a larger $c$ in (b), $y(c) > 1$ is the solution of the stationary condition which is approximated to that of Eq.~\eqref{eq:VY6Y4p-minimum-inf}.
    A large $y(c) > 1$ in the table (c) for $c\gtrsim 2$ is approximately given by the solution of Eq.~\eqref{eq:minimum-44p}.
	}
	\label{tab:potential-summary}
\end{table}

Before moving on to a concrete analysis, let us summarize our results here.
Tab.~\ref{tab:potential-summary} summarizes where the vacua of the potentials exist for each value of $c$ as it varies.
The potential minimum is shifted from the vacuum obtained only with one modular form to a new vacuum due to a $c$-dependent contribution of the additional modular form.
A non-trivial vacuum away from the fixed points is realized for the scalar potential with $Y^{(4)}_{\bm{1}'}$ for a large $c$.
After all, the vacua obtained are CP-conserving ones with the positive cosmological constant.
The $X$ fields have positive mass squared at these vacua as shown in Sec.~\ref{sec:X-mass}.

\subsection{Vacuum of $V_{6 + 4}$}
\label{sec:V-6+4}

\begin{figure}[t]
    \centering
        \begin{subfigure}{0.31\textwidth}
            \centering
            \includegraphics[width=\textwidth]{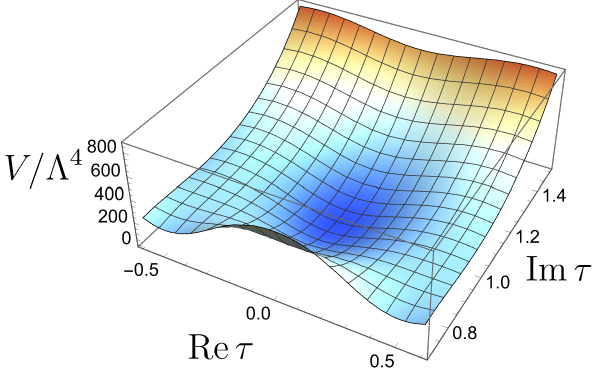}
            \subcaption{$c=1$}
        \end{subfigure}
        \begin{subfigure}{0.31\textwidth}
            \centering
            \includegraphics[width=\textwidth]{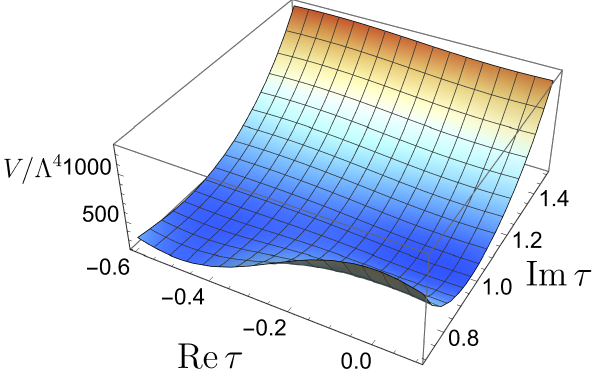}
            \subcaption{$c=2.5$}
        \end{subfigure}
        \begin{subfigure}{0.31\textwidth}
            \centering
            \includegraphics[width=\textwidth]{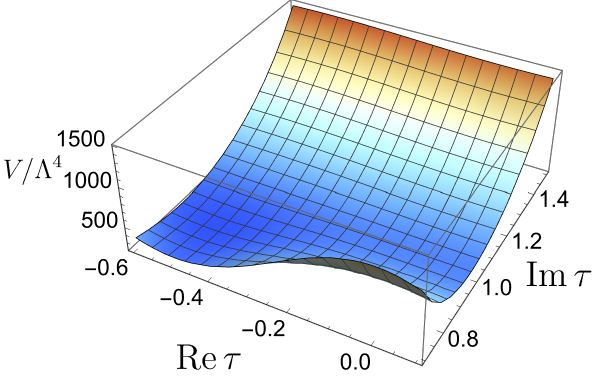}
            \subcaption{$c=5$}
        \end{subfigure}
    \caption{
        The 3D plots of $V_{6+4}$ for $c=1,\ 2.5,$ and $5$, respectively.
        The potential minimum is at $ \tau = \iu$ for a small $c$, while it moves to $\tau = \omega$ as $c$ increases.
        This is because $V_{6+4} \sim V^{(4)}_{\bm{1}}$ for a large $c$.
    }
    \label{fig:VY6Y4}
\end{figure}

Let us start our discussion with focusing on the vacuum of $V_{6+4}$.
For details, see Apps.~\ref{sec:details-of-potential} and \ref{sec:Hessian-analysis}.
Fig.~\ref{fig:VY6Y4} shows that the minimum of $V_{6+4}$ shifts from $\tau = \iu$ to $\tau = \omega$ as $c$ increases. 
A critical change occurs when $c \gtrsim 2$ as discussed in the following.
In the following part, we study how the potential minimum changes in a perturbative way around the fixed points, and it turns out that the vacua at the fixed points are stable against a perturbation, depending on $c$.

\paragraph{Expansion around $\tau = \iu$.}

Let us expand the scalar potential $V_{6 +4}$ by using $\tau \eqqcolon \iu + x + \iu y$:
\begin{align}
	V_{6 +4} (x,y) & \approx \bigl[ 2.84\,y^2 - 2.84\,y^3 + 8.44\,y^4 
		\nn \\
            &\quad +\left(
                2.84-2.84\,y^2-7.09\,y^4
            \right)x^2
            -15.5\,x^4  \bigr] \times 10^3
            \nn \\
            & \quad + c^2 \bigl[ 33.9
            + 237\, y^2 
            - 237\, y^3
            + 622\, y^4 
            \nn \\
            &\quad
            + \left(
                - 305 
                + 846\,y 
                - 1.10\times10^3\,y^2 \right) \, x^2
            + 1.20\times10^3 \, x^4 \bigr] .
\end{align}
Here,  we expand the potential to the fourth order in fluctuations to see the $c$-dependence of the VEV of $\tau$.
The first derivatives of $V_{6 +4}$ are given by 
\begin{align}
    \frac{\del V_{6+4}}{\del x} &= 4 \mc{X}_4(c) \Bigl(
        x^2 + \frac{1}{2} \frac{\mc{X}_2 (c,y)}{\mc{X}_4(c)}
    \Bigr) x,
    \label{eq:delVr/delx}
    \\
    \frac{\del V_{6+4}}{\del y}&= \mc{Y}_1(c,x) + \mc{Y}_{2}(c,x) y + \mc{Y}_3(c) y^2 + \mc{Y}_4(c) y^3.
    \label{eq:delVr/dely}
\end{align}
The coefficients $\mc{X}_{2,4}$ and $\mc{Y}_{1,2,3,4}$ are summarized in Tabs.~\ref{tab:potential_i_expand4_xdiff_coefficient-app} and \ref{tab:potential_i_expand4_ydiff_coefficient-app} in App.~\ref{sec:details-of-potential}.
For small $c$ and $y \ll 1$, the signatures of $\mc{X}_2$ and $\mc{X}_4$ become 
\begin{align}
    \mc{X}_2 \approx 823.23 - 2755.7 y - 14367 y^2 >0,
    \qquad 
    \mc{X}_4 \approx (1.12 c^2 - 15.5) \times 10^{3} < 0.
\end{align}
Then $V_{6+4}$ has the minimum at $x = 0$.
At $x=0$, we obtain $\mc{Y}_1 = 0$ and the stationary condition of Eq.~\eqref{eq:delVr/dely} reduces to
\begin{align}
    \frac{\del V_{6+4}}{\del y} \bigg|_{x=0} = \bigl[ \mc{Y}_4 (c) y^2 + \mc{Y}_3(c) y + \mc{Y}_2(x =0,c) \bigr] y=0.
    \label{eq:delV-6+4/y-x=0}
\end{align}
We find that the discriminant of $y$'s second order equation in the bracket of Eq.~\eqref{eq:delV-6+4/y-x=0} is always negative 
\begin{align}
    \mc{Y}_3^2(c) - 4 \mc{Y}_2(x=0,c) \mc{Y}_4(c) < 0,
\end{align}
and hence the stationary point is given by $y=0$.
Thus, the fixed point $\tau = \iu$ remains vacuum for a small $c$.
At $\tau = \iu$ vacuum, $Y^{(6)}_{\bm 1} =0$ and $Y^{(4)}_{\bm 1} \neq 0$ and hence $\braket{V_{6+4}}  \sim c^2 \braket{|Y^{(4)}_{\bm 1}|^2} > 0$.

The threshold value of $c$ that triggers the vacuum shift from $\tau = \iu$ to $\tau = \omega$ is discussed with the expansion around $\tau = \omega$ as below.

\begin{figure}[t]
	\centering
	\includegraphics[width=0.5\textwidth]{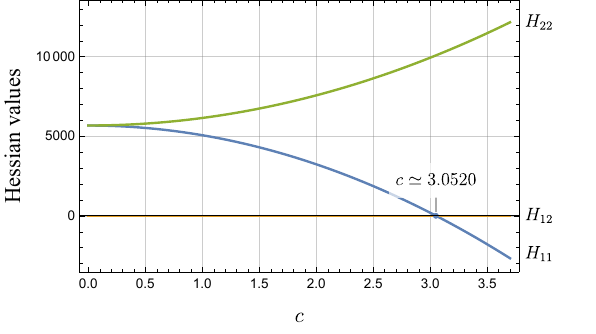}
	\caption{The $c$-dependence of the Hessian matrix at the $\tau = \iu$ vacuum.
	$H_{11}$ becomes negative when $c \gtrsim 3$ and $H_{22}$ monotonically increases.}
	\label{fig:f_VY6Y4_i_Hessian}
\end{figure}

In order to see the stability of the vacuum at $\tau = \iu$, let us introduce the Hessian matrix: 
\begin{align}
    \begin{pmatrix}
        H_{11}(x,y;c) & H_{12} (c,x,y)
        \\
        H_{21}(x,y;c) & H_{22}(c,x,y)
    \end{pmatrix}
	&\coloneqq
	\begin{pmatrix}
     \del^2 V/\del x^2 & \del^2 V/ \del x \del y \\
        \del^2 V/\del y \del x & \del^2 V/ \del y^2
	\end{pmatrix}
	\label{eq:hessian-matrix}
        \\
	&= \begin{pmatrix}
		\del^2 V / \del x^2 & 0 \\
		0 & \del^2 V / \del y^2
	\end{pmatrix}.
\end{align}
In the last equality, we used the fact that the off-diagonal components vanish when CP is conserved at the vacuum.
Here, the CP transformation is given by
\begin{align}
    \mr{CP}: \tau \mapsto -\bar{\tau} .
\end{align}
Thus, the vacuum at $\tau = \iu$ preserves CP.
The $c$-dependence of the Hessian matrix at $\tau = \iu$ is shown in Fig.~\ref{fig:f_VY6Y4_i_Hessian}.
Recovering the overall scale of $\Lambda$,
we obtain the modulus mass at $\tau = \iu$
\begin{align}
m_\tau \sim \Lambda^2 |\partial_\tau Y^{(6)}_{\bm 1}(\tau = i)|
\sim 10\Lambda^2 .
\label{eq:tau-mass-i}
\end{align}
Note that $H_{11}$ becomes tachyonic when $c \gtrsim 3$. 
Then, the vacuum at $\tau = \iu$ becomes unstable since $Y^{(4)}_{\bm{1}} \neq 0$ at $\tau =\iu$ and $V_{6+4}$ is approximately given by $c^2 |Y^{(4)}_{\bm{1}}|^2$ which costs energy
for such a large $c$.

\paragraph{Expansion around $\omega$.}

Using Eqs.~\eqref{eq:Y-expansion-omega-41}, \eqref{eq:Y-expansion-omega-61}, and $\tau \eqqcolon \omega + x + \iu y$, the potential can be expanded as
\begin{align}
	V_{6+4} &\approx 
    \left(224 -448 y^2 +7.51\times10^3 y^3-1.22\times10^4 y^4\right) 
	\nn\\
        &\quad 
        +\left(
            -448 -2.05\times10^4 y+7.32\times10^4 y^2
        \right)x^2  
        -1.16\times10^4 x^4
        \nn \\
        & \quad 
        + c^2 
        \Bigl[
            328y^2 -379y^3 -109y^4
            +\left(
                328 -379y-656y^2
            \right)\,x^2 
            -546x^4
        \Bigr] .
\end{align}
The first derivatives of the potential are given by 
\begin{align}
	\frac{\del V_{6+4}}{\del x}
        \approx 2\big\{
                -449 + 328 c^2
            \big\} x,
            \qquad 
	 \frac{\del V_{6+4}}{\del y}
        \approx 2\bigl\{
            -449 + 328 c^2 
        \bigr\}y.
\end{align}
For $c \gtrsim 1$, the second term in the bracket of two $\partial V$'s becomes dominant and $\tau = \omega$ becomes a local minimum: $\del^2 V_{6+4}/\del x^2 = \del^2 V_{6+4}/\del y^2 >0$.
For $c \gtrsim 2$, we numerically find $V_{6+4}(\tau = \iu) \sim c^2 |Y^{(4)}_{\bm 1}(\tau = \iu)|^2 \geq V_{6+4} (\tau = \omega) \sim |Y^{(6)}_{\bm 1}(\tau = \omega)|^2$, and hence $\tau = \omega$ becomes the global minimum.
Recovering the overall scale of $\Lambda$,
we obtain the modulus mass at $\tau = \omega$
\begin{align}
    m_\tau \sim c\Lambda^2 |\partial_\tau Y^{(4)}_{\bm 1}(\tau = \omega)| \sim 10 c \Lambda^2 .
\end{align}
Note that $\tau = \omega$ is the CP-conserving minimum with $T$ symmetry of $\tau \equiv \tau -1$:
\begin{align}
    \mr{CP}:~ \tau= \omega = -\frac{1}{2} + \iu \frac{\sqrt{3}}{2}
    \mapsto \frac{1}{2} + \iu \frac{\sqrt{3}}{2} \equiv -\frac{1}{2} + \iu \frac{\sqrt{3}}{2} = \omega.
\end{align}
At $\tau = \omega$ vacuum, $Y^{(6)}_{\bm 1} \neq 0$ and $Y^{(4)}_{\bm 1} = 0$ and hence $\braket{V_{6+4}} \sim \braket{|Y^{(6)}_{\bm 1}|^2} > 0$.

\subsection{Vacuum of $V_{6+ 4'}$}
\label{sec:V-6+4'}

\begin{figure}
    \centering
    \begin{subfigure}{0.31\textwidth}
            \centering
            \includegraphics[width=\textwidth]{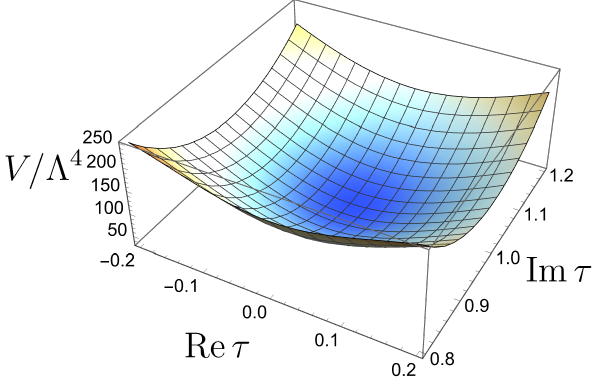}
            \subcaption{$c=1$}
    \end{subfigure}
        \begin{subfigure}{0.31\textwidth}
            \centering
            \includegraphics[width=\textwidth]{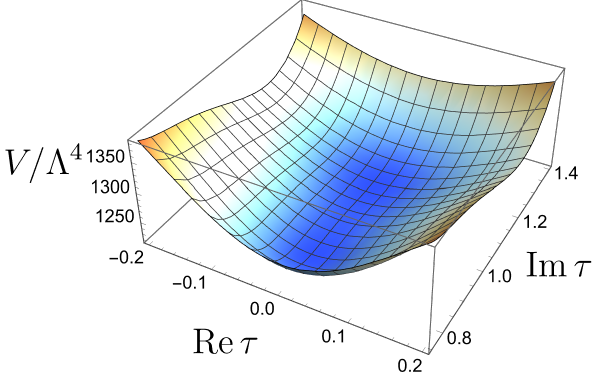}
            \subcaption{$c=6$}
        \end{subfigure}
        \begin{subfigure}{0.31\textwidth}
            \centering
            \includegraphics[width=\textwidth]{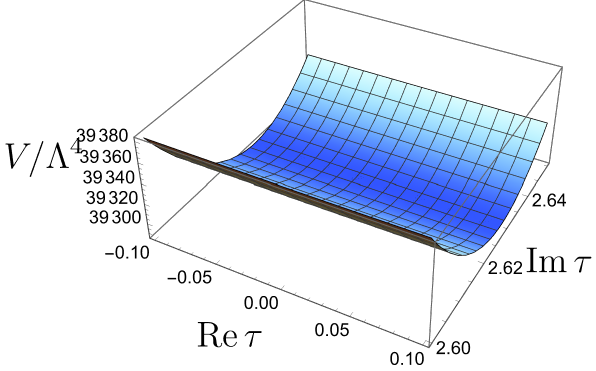}
            \subcaption{$c=100$}
        \end{subfigure}
    \caption{
    The 3D plots of $V_{6+4'}$ for $c = 1, \ 6$, and $100$, respectively.
     $V_{6+4'}$ can have a vacuum at $\braket{\tau} \to \iu \infty$ as $c$ increases. 
    For $c = \mc{O}(100)$, $V_{6+4'}$ has the minimum at $ \Im\tau \sim 3$.
    }
    \label{fig:VY6Y4p}
\end{figure}

Next, we consider the vacuum of the potential $V_{6+4'}$.
Fig.~\ref{fig:VY6Y4p} shows that the potential can shift away from the fixed point $\tau = \iu$ to $\Im \tau \gg 1$ as $c$ increases.
In the following part, let us show this change of $\braket{\tau}$ in Fig.~\ref{fig:VY6Y4p} by perturbative analysis.

\paragraph{Expansion around $\tau = \iu$.}

As in the case of $V_{6+4}$, we expand the potential to the fourth order in fluctuations.
\begin{align}
	V_{6+4'}(x, y) &\approx \bigl[ 2.84\,y^2 - 2.84\,y^3 + 8.44\,y^4 
		 \nn \\
            &\quad +\left(
                2.84-2.84\,y -7.09\,y^2
            \right)x^2
            -15.5\,x^4 \bigr] \times 10^3
		\nn \\
		& \quad + c^2 \bigl[
			33.9
            - 88.0\, y^2 
            + 88.0\, y^3
            + 11.6\, y^4 
            \nn \\
            &\quad
            + \left(
                20.2 
                + 128\,y 
                + 360\,y^2 \right) \, x^2
            - 62.7 \, x^4
		\bigr],
\end{align}
where we use $\tau \eqqcolon \iu + x + \iu y$.
The first derivatives of $V_{6+4'}$ are similarly given by Eqs.~\eqref{eq:delVr/delx} and \eqref{eq:delVr/dely}, and see Tabs.~\ref{tab:potential_i_expand4_xdiff_coefficient-app} and \ref{tab:potential_i_expand4_ydiff_coefficient-app} in App.~\ref{sec:details-of-potential} for the details of $\mc{X}$ and $\mc{Y}$.
For $y \ll 1$, we focus on a small $c$ and then find
\begin{align}
    \mc{X}_2 &= 2836.9 - 2836.9 y - 7092.2 y^2 + c^2(20.224 + 128.49 y + 360.09 y^2) > 0, \\
        \mc{X}_4 &= -62.735 c^2 
        -15532<0.
\end{align}
Then $\del V_{6+4'}/ \del x =0$ is satisfied by $x = 0,\ \pm \sqrt{- \mc{X}_2/(2 \mc{X}_4)}$, and $x=0$ is the minimum.

\begin{figure}[t] 
    \centering
    \includegraphics[width=0.45\textwidth]{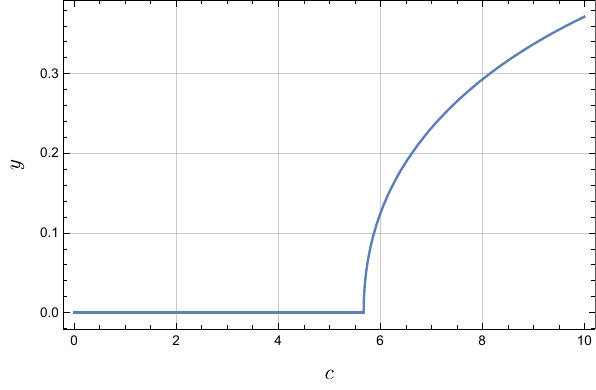}
    \quad 
    \includegraphics[width=0.45\textwidth]{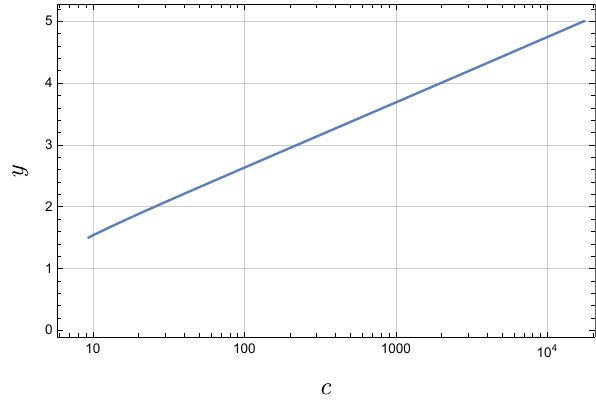}
    \caption{
        (Left)
        The behavior of the non-zero solution in Eq.~\eqref{eq:yofc-V6+4p-taui-2} as the function of $c$. The vacuum shifts in the $\Im \tau$ direction as $c$ increases.
        (Right)
        The behavior of the solution $y$ as the function of $c$, which is approximately given by Eq.~\eqref{eq:VY6Y4p-minimum-inf}.
    It is found that $y \sim 2.5$ is realized with about $c \sim 100$~\cite{Petcov:2023vws,Abe:2023qmr}.
    }
    \label{fig:VY6Y4p_minimum.pdf}
\end{figure}

Let us consider the stationary condition of the imaginary direction:
\begin{align}
    \frac{\del V_{6+4'}}{\del y}\bigg|_{x =0} =\bigl(
        \mc{Y}_4(c) y^2 + \mc{Y}_3(c) y + \mc{Y}_2(x=0,c)
    \bigr) y = 0.
    \label{eq:delV-6+4'/dely-x=0}
\end{align}
Whether the solution of $y \neq 0$ exists or not depends on the discriminant of the equation in the brackets of \eqref{eq:delV-6+4'/dely-x=0}.
It is found that numerically
\begin{align}
    \mc{Y}_3^2 - 4 \mc{Y}_2 \mc{Y}_4 > 0 
    \quad 
    \text{when}\ 
    c \gtrsim 5,
\end{align}
and hence the stationary points of $V_{6 + 4'}$ are
\begin{align}
    y(c) = \begin{cases}
        \displaystyle
        0 & c \lesssim  5
        \\
        \displaystyle
        \frac{- \mc{Y}_3 + \sqrt{\mc{Y}_3^2 - 4 \mc{Y}_2 \mc{Y}_4}}{2 \mc{Y}_4},~0 & c \gtrsim  5
    \end{cases},
    \label{eq:yofc-V6+4p-taui}
\end{align}
where we drop $\displaystyle \frac{- \mc{Y}_3 - \sqrt{\mc{Y}_3^2 - 4 \mc{Y}_2 \mc{Y}_4}}{2 \mc{Y}_4}$, which is located outside the fundamental domain. For $c \gtrsim  5$, $y=0$ is the local maximum.
The explicit form of the non-zero solution is given by
\begin{align}
    &\frac{- \mc{Y}_3 + \sqrt{\mc{Y}_3^2 - 4 \mc{Y}_2 \mc{Y}_4}}{2 \mc{Y}_4}
    \nn \\
    &= 
    -2.8413 
    +
    \frac{2155.3 +4.7426 \sqrt{-3569.4+93.7802 c^2+0.52736 c^4}}{726.34 + c^2}  .
    \label{eq:yofc-V6+4p-taui-2}
\end{align}
The left panel of Fig.~\ref{fig:VY6Y4p_minimum.pdf} shows the behavior of non-zero $y(c)$ as a function of $c$.
The VEV of $y$ stays zero for $c \leq 5.68$, and increases monotonically when $c \gtrsim 5$.

\begin{figure}[t]
    \centering
    \includegraphics[width=0.48\textwidth]{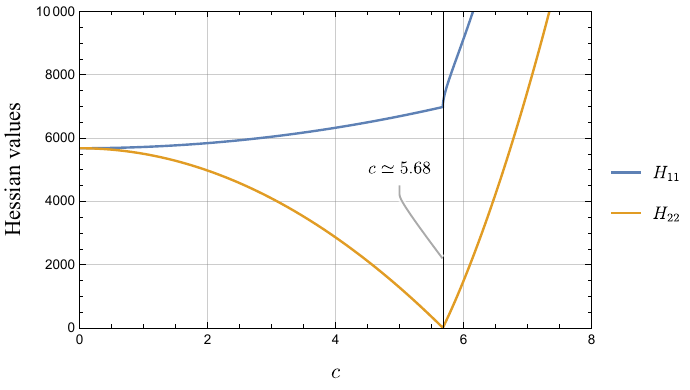}
    \quad 
    \includegraphics[width=0.48\textwidth]{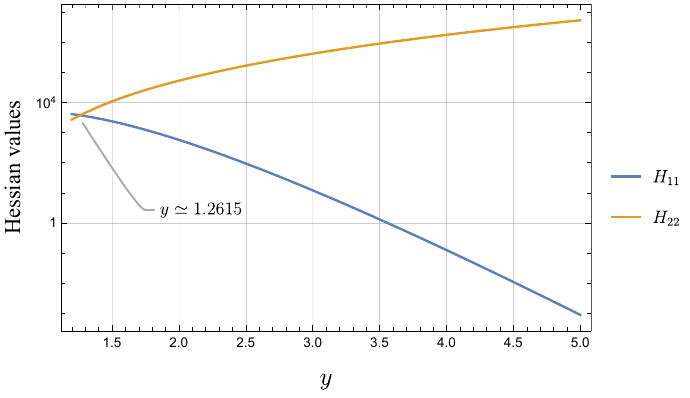}
    \caption{
		(Left)
		The $c$-dependence of the Hessian diagonal components.
		(Right)
		The values of the Hessian matrix evaluated at $\tau = \iu y(c)$.
    See the text for the details.}
    \label{fig:VY6Y4p_Hessianvalues}
\end{figure}

The $c$-dependence of the Hessian diagonal components \eqref{eq:hessian-matrix} is shown in Fig.~\ref{fig:VY6Y4p_Hessianvalues}. 
Note that the off-diagonal components of the Hessian are vanishing at the CP-conserving vacuum at $\tau = i(1 +y(c))$.
Recovering the overall scale of $\Lambda$, we obtain the modulus mass at $\langle \tau \rangle = i(1 +y(c))$
\begin{align}
    m_\tau \sim \Lambda^2 |\partial_\tau Y^{(6)}_{\bm 1}(\langle \tau \rangle)|
    \sim 10\Lambda^2 .
    \label{eq:tau-mass-iy}
\end{align}
The mass difference between $x$ and $y$ gets bigger as $c$ increases, and the reason is shown later at the end of this section.
In the left panel the Hessian is evaluated at $x = y =0$ for $c \lesssim 5.68$, whereas it is evaluated at $\displaystyle x =0,~ y(c)= \frac{- \mc{Y}_3 + \sqrt{\mc{Y}_3^2 -4 \mc{Y}_2 \mc{Y}_4}}{2 \mc{Y}_4}$ for $c \gtrsim 5.68$.
Then, the masses of the modulus fields become monotonically heavy as $c$ increases.

\paragraph{Expansion around $\tau = \iu \infty$.}

The contribution of the runaway potential $V^{(4)}_{\bm{1}'} \sim |Y^{(4)}_{\bm{1}'} |^2$ is dominant in $V_{6+4'}$ for a large $c$, and the vacuum exists at a large $\Im \tau$.
For a large $\Im \tau$ VEV, the scalar potential is given by following form with a $q$-expansion
\begin{align}
    V_{6+4'}&=(2y)^6 |1 - 504 q|^2 + c^2 (2y)^4 |-12q^{1/3}|^2 + \mc{O}(q^3)
    \nn \\
    &= (2y)^6 (1 - 1008 \ee^{-2 \pi y} \cos 2 \pi x + 504^2 \ee^{-4 \pi y}) + c^2 (2y)^4 144 \ee^{- 4 \pi y/3} + \mc{O}(q^3).
    \label{eq:V6+4p-expansion}
\end{align}
Here, we used $q=\exp(2\pi \iu \tau) = \exp[2\pi (\iu x -y)]$ and $y \gg 1$.
The first derivative of $V_{6+4'}$ with respect to $x$ is
\begin{align}
    \frac{\del V_{6+4'}}{\del x} \propto + \sin 2 \pi x.
\end{align}
Thus, the minimum is located at $x = 0$ independently of $c$. Hence the vacuum is still CP symmetric for a large $c$.
The potential gradient along the $y$-direction at $x =0$ is
\begin{align}
    \frac{\del V_{6+4'}}{\del y}\bigg|_{x=0}
    &\approx 384 y^3 \Bigl[
        \ee^{-4 \pi y} ( \ee^{2\pi y} - 504) (336 \pi y + \ee^{2\pi y} - 504) y^2 - 8 c^2 \ee^{- 4 \pi y/3} (\pi y -3)
    \Bigr].
\end{align}
In the $y \gg 1$ limit, the stationary condition at the leading order in $q$ reduces to
\begin{align}
    \frac{\del V_{6+4'}}{\del y}\bigg|_{x=0} \sim 384 y^3 \bigl[
        -8 c^2 \ee^{-4 \pi y/3} ( \pi y-3) + y^2
    \bigr] =0.
\end{align}
Equivalently, the above condition is rewritten as
\begin{align}
    y + \frac{3}{4\pi} \log \frac{y^2}{y- 3/\pi} = \frac{3}{2\pi}\log c + \frac{3}{4\pi}\log 8\pi.
    \label{eq:VY6Y4p-minimum-inf}
\end{align}
The right panel in Fig.~\ref{fig:VY6Y4p_minimum.pdf} shows the solution $y(c)$ of this equation as a function of $c$. 
$y(c)$ turns out to increase as $c$ does.
In order to see the stability around the stationary point at $\tau = \iu y(c)$, the second derivative of the potential is evaluated as 
\begin{align}
    \frac{\del^2 V_{6+4'}}{\del y^2}\bigg|_{x=0,~y=y(c)} &= 384 \ee^{- 4 \pi y} y^4 \Bigl[
        -8c^2 \ee^{8 \pi y/3}(\pi y -3) + (\ee^{2\pi y} - 504) (336 \pi y + \ee^{2\pi y} - 504 ) y^2
    \Bigr]\bigg|_{y=y(c)}
    \nn \\
    &\sim 128y(c)^4 \frac{4\pi^2 y(c)^2 - 9\pi y(c) -18}{\pi y(c) -3} .
	\label{eq:del2V-6+4'dely2}
\end{align}
The $y(c)$-dependence of the diagonal components of the Hessian is shown in the right panel in Fig.~\ref{fig:VY6Y4p_Hessianvalues}, while the off-diagonal components are vanishing. 
The mass of $\Im \tau$ becomes heavier as $y(c)$ (and $c$) increases.
This is because the mass is proportional to Eq.~\eqref{eq:del2V-6+4'dely2}.
On the other hand, the mass of $\Re \tau$ gets lighter as $y(c)$ increases. 
This is seen from $V_{6+4'} \sim -\ee^{-2 \pi y} \cos 2\pi x$.
Hence, recovering the overall scale of $\Lambda$, we roughly find
\begin{align}
    m_x \sim 10^2 y(c)^4 e^{-\pi y(c)} \Lambda^2 ,
    \qquad
    m_y \sim 10^2 y(c)^{7/2} \Lambda^2 ,   
\end{align}
where we have canonically normalized the masses by multiplying $y(c)^2$ by the second derivatives of $V_{6+4'}$.
The explicit form of the Hessian matrix component is shown in Eq.~\eqref{eq:H11-V64p-iinf}.

In this model, the vacuum exists at a point apart from any fixed points.
Thus, $\braket{V_{6+4'}} \sim \braket{|Y_{\bm 1}^{(6)}|^2 + |Y_{{\bm 1}'}^{(4)}|^2} > 0$.

\subsection{Vacuum of $V_{4+4'}$}

\begin{figure}[t]
    \centering
    \begin{subfigure}{0.31\textwidth}
        \centering
        \includegraphics[width=\textwidth]{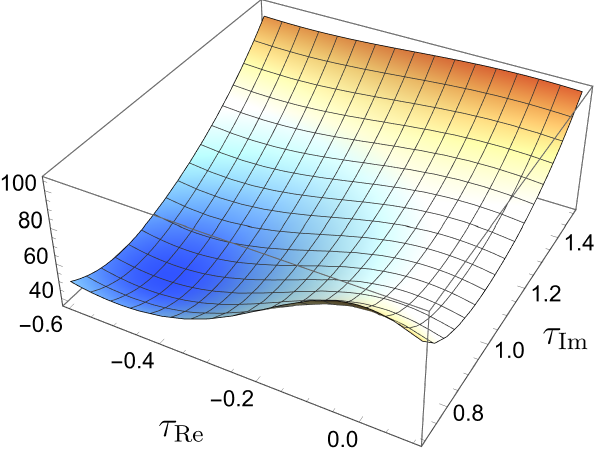}
        \caption{$c=1$}
    \end{subfigure}
    \begin{subfigure}{0.31\textwidth}
        \centering
        \includegraphics[width=\textwidth]{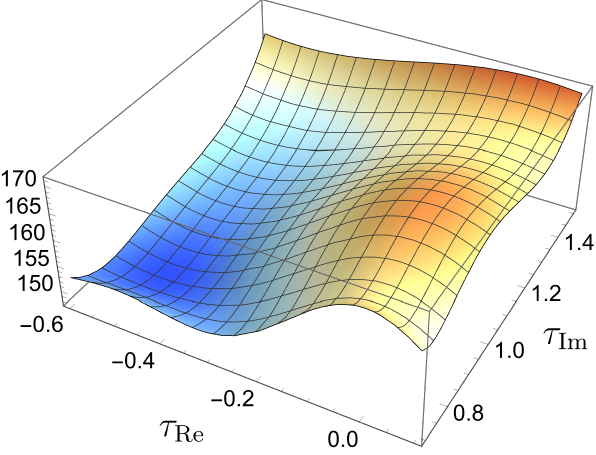}
        \caption{$c=2$}
    \end{subfigure}
        \begin{subfigure}{0.31\textwidth}
        \centering
        \includegraphics[width=\textwidth]{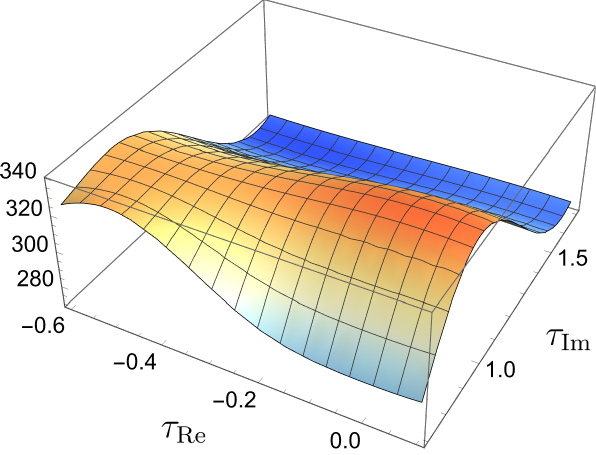}
        \caption{$c=3$}
    \end{subfigure}
    \caption{
    The 3D plots of $V_{4+4'}$ for $c = 1,\ 2,$ and $3$, respectively.
    }
    \label{fig:VY4Y4p}
\end{figure}

Similarly to $V_{6+4'}$, the potential $V_{4+4'}$ can have the minimum at $\tau = -1/2 + \iu \Im \tau$, where $\Im \tau \geq \sqrt{3}/2$, depending on $c$ as shown in Fig.~\ref{fig:VY4Y4p}.
We show $c$-dependence of the modulus VEV as in the previous sections.

\paragraph{Expansion around $\tau = \omega$.}
\label{sec:V-4+4'}

Eqs.~\eqref{eq:Y-expansion-omega-41} and \eqref{eq:Y-expansion-omega-61} give us the following potential expansion:
\begin{align}
        V_{4+4'}(x,y)
        =&~
        328y^2-379 y^3-109 y^4 
        +\left(328-379y-656y^2\right)\,x^2 
        -546x^4 
        \nn \\
        &+c^2
        \left\{
            36.9-49.2 y^2-351 y^3+485 y^4
            +\left(-49.2+940 y+3.04\times10^3 y^2\right)x^2
            +551 x^4
        \right\},
\end{align}
where $\tau \eqqcolon \omega + x + \iu y$.
The first derivatives of $V_{4+4'}$ are given by 
\begin{align}
	\frac{\del V_{4+4'}}{\del x} &= 2 (327.6 - 49.16 c^2) x,
	\\
	\frac{\del V_{4+4'}}{\del y} &= 2 (327.6 - 49.16 c^2) y.
\end{align}
Here, the only leading contribution is considered.
For a small $c \lesssim 2$, the minimum is located at $x = y = 0$:
the second derivatives of $V_{4+4'}$ are given by
\begin{align}
	\frac{\del^2 V_{4+4'}}{\del x^2} \bigg|_{x= y=0}= \frac{\del^2 V_{4+4'}}{\del y^2} \bigg|_{x= y=0} = 2 (328 - 49.2 c^2),
	\qquad
	\frac{\del^2 V_{4+4'}}{\del x \del y} = 0.
\end{align}
Thus, recovering the overall scale of $\Lambda$, we obtain the modulus mass at $\tau  = \omega$
\begin{align}
    m_\tau \sim \Lambda^2 |\partial_\tau Y^{(4)}_{\bm 1}(\tau =\omega)|
    \sim 10\Lambda^2 .
\end{align}
When $c$ gets larger than $2$ and hence the contribution from $Y^{(4)}_{\bm{1}'}$ becomes dominant
in $V_{4+4'}$, tachyonic modes appear at $x=y=0$.
The instability indicates that the vacuum can be shifted from $\tau = \omega = -1/2 + \iu \sqrt{3}/2$ to a larger $\Im\tau$.

\paragraph{Expansion around $\tau = -1/2 + \iu \infty$.}

\begin{figure}[t]
	\centering
	\includegraphics[width=0.41\textwidth]{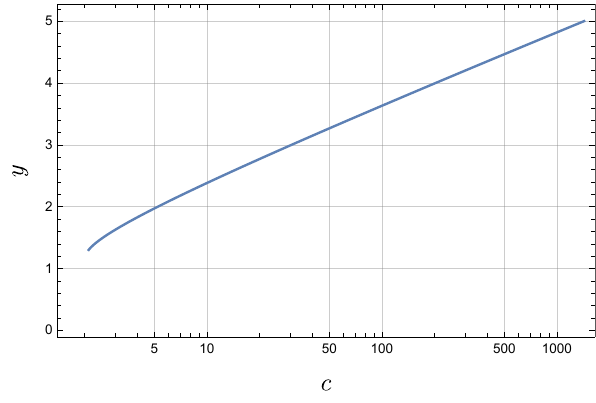}
	\qquad 
	\includegraphics[width=0.48\textwidth]{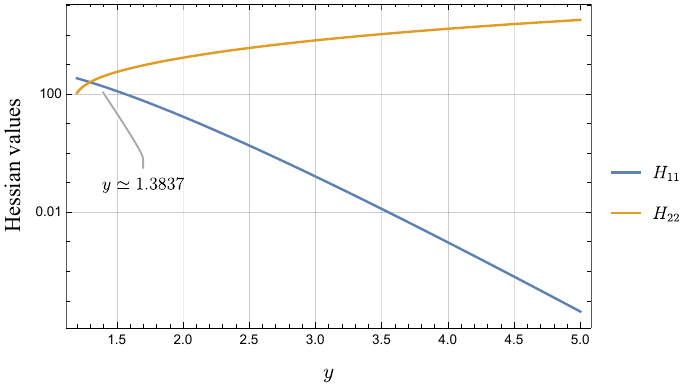}
	\caption{
		(Left) The behavior of the solution $y$ as the function of $c$ which is approximately given by Eq.~\eqref{eq:minimum-44p}.
		(Right) The diagonal components of the Hessian matrix at $\tau = \iu y$, where $y$ is approximately given by the solution of Eq.~\eqref{eq:minimum-44p}.
	}
	\label{fig:Y4Y4p-plot}
\end{figure}

When a $c$ is large, we use the $q$-expansion and the modular forms are expressed as 
\begin{align}
	Y^{(4)}_{\bm{1}}  \approx 1 + 240 q,
	\qquad 
	Y^{(4)}_{\bm{1}'} \approx - 12 q^{1/3},
\end{align}
for $|q| \ll 1$, where $q= e^{2\pi \iu \tau}$.
We parameterize $\tau = x + \iu y~(y \gg 1)$ and the potential is given by
\begin{align}
	V_{4 + 4'} &\approx (2y)^4 \left(
		|1 + 240 q|^2 + c^2 \left|- 12 q^{1/3} \right|^2
	\right)
	\nn \\
	&=16 y^4
	\Bigl[
		1 + 57600 \ee^{- 4 \pi y} + 480 \ee^{-2\pi y} \cos (2 \pi x) + 144 c^2 \ee^{- 4\pi y/3}	
	\Bigr].
\end{align}
We note that the signature of $\cos (2\pi x)$ is opposite to that in Eq.~\eqref{eq:V6+4p-expansion}.
Due to this cosine function, this potential has the minimum at $x = -1/2$:
\begin{align}
	\frac{\del V_{4 + 4'}}{\del x} \propto - \sin (2\pi x) , \qquad
 \frac{\del^2 V_{4 + 4'}}{\del x^2} \propto - \cos (2\pi x). 
\end{align}
Hence, CP is preserved at the vacuum with the $T$-symmetry like in a case of $\tau = \omega$.
The potential gradient with respect to $y$ is
\begin{align}
	\frac{\del V_{4+4'}}{\del y}\bigg|_{x=-1/2} &= 64 y^3 \Bigl[
		(1 - 240 \ee^{-2 \pi y} )(1 - 240 \ee^{-2\pi y} + 240 \pi \ee^{-2 \pi y}y)
		- 48 c^2 \ee^{- 4\pi y/3}(\pi y -3)
	\Bigr].
\end{align}
For $ y \gg 1$, the stationary condition approximately reduces to
\begin{align}
	\frac{\del V_{4+ 4'}}{\del y}\bigg|_{x=-1/2} \approx 64 y^3 \Bigl[
		1 - 48 c^2 \ee^{-4 \pi y/3} (\pi y - 3)
	\Bigr] = 0.
\end{align}
Equivalently, the above condition reads
\begin{align}
	1 - 48 c^2 \ee^{-4 \pi y/3} ( \pi y- 3) =0.
	\label{eq:minimum-44p}
\end{align}
The solution $y(c)$ of this Eq.~\eqref{eq:minimum-44p} as a function of $c$ is shown in the left panel in Fig.~\ref{fig:Y4Y4p-plot}. The $y(c)$ gets larger as $c$ increases.
The second derivative of $V_{4 + 4'}$ is evaluated as
\begin{align}
	\frac{\del^2 V_{4+4'}}{\del y^2}\bigg|_{x=-1/2,y=y(c)} &= 64 \ee^{-4 \pi y} y^2 \Bigl[
		16 c^2 \ee^{8 \pi y/3} (4 \pi^2 y^2- 24 \pi y + 27)
		- 480 \ee^{2\pi y}(\pi^2 y^2 -4 \pi y +3) 
		\nn \\
		& \qquad \qquad \qquad
		+57600 (4 \pi^2 y^2 - 8 \pi y + 3) + 3 \ee^{4 \pi y}
	\Bigr]\bigg|_{y=y(c)}
	\nn \\
	& \approx 64 y(c)^2 \Bigl[
		16 c^2 \ee^{- 4 \pi y(c)/3} ( 4 \pi^2 y(c)^2 - 24 \pi y(c) + 27)  + 3
	\Bigr].
\end{align}
In the last equality, we picked up just the leading terms in $|q|\ll 1$.
Putting the solution of Eq.~\eqref{eq:minimum-44p} into this equation, we obtain the following simple form:
\begin{align}
	\frac{\del^2 V_{4+4'}}{\del y^2} \bigg|_{\text{solution of Eq.~\eqref{eq:minimum-44p}}}
	 = \frac{64 \pi}{3} \frac{y(c)^3 ( 4 \pi y(c) - 15)}{\pi y(c) -3},
\end{align}
and the signature of this equation is positive if $y(c) \geq 15 / 4\pi$ for $c \gtrsim 2$.
The $y(c)$ is the minimum of $V_{4+4'}$ for a large $y$ and thus $x=-1/2$ and $y > 15 / 4\pi$ is the minimum for $c \gtrsim 2$.

We give some comments on the Hessian analysis.
We note that the off-diagonal component with $x = -1/2$ vanishes and the $y$-dependence of the diagonal components is shown in the right panel in Fig.~\ref{fig:Y4Y4p-plot}.
This figure shows numerically that the vacuum at $\langle \tau \rangle = -1/2 + \iu y(c)$ is stable.
Hence, recovering the overall scale of $\Lambda$, we roughly find
\begin{align}
    m_x \sim 10^2 y(c)^3 e^{-\pi y(c)} \Lambda^2 ,
    \qquad
    m_y \sim 10^2 y(c)^{5/2} \Lambda^2 ,   
\end{align}
where we have canonically normalized masses by multiplying $y(c)^2$ by the second derivatives of $V_{6+4'}$ as in the previous section.

In this model, there exists the vacuum at a point apart from any fixed points.
Thus, $\braket{V_{4+4'} } \sim \braket{|Y_{\bm 1}^{(4)}|^2 + |Y_{{\bm 1}'}^{(4)}|^2} > 0$.

\subsection{$X$ masses}
\label{sec:X-mass}

We numerically show the mass eigenvalues of $X$ fields, $m_{\pm}$ given in Eq.~(\ref{eq:X-mass-eigen}).
With a recovered overall scale of $\Lambda$, the explicit forms read  
\begin{align}
    m^2_X 
    &=  \Lambda^4 (2 \Im\tau)^{-h}  
        \biggl[
            \frac{|Y_X(\tau)|^2}{\Lambda_X^2} 
            + c^2 |Y_\Xi(\tau)|^2 (2 \Im\tau)^{k_X - k_\Xi} 
            + \frac{|(h+ k_X)Y_X(\tau) - 2 \iu \Im \tau \del_\tau Y_X(\tau)|^2}{h}
        \biggr], \\ 
    m^2_\Xi 
    &= \Lambda^4 (2 \Im\tau)^{-h}  
        \biggl[
            c^2 \frac{|Y_\Xi(\tau)|^2}{\Lambda_\Xi^2} 
            + |Y_X(\tau)|^2 (2 \Im\tau)^{-k_X + k_\Xi} 
            + c^2 \frac{|(h+ k_\Xi)Y_\Xi(\tau) - 2 \iu \Im \tau \del_\tau Y_\Xi(\tau)|^2}{h}
        \biggr], \\
    M^2
    &= c \Lambda^4 \frac{(2 \Im\tau)^{-h}}{h} 
    \nn \\ 
        &\quad \times 
        \biggl[
        \Bigl\{
            \bigl( (h+ k_X) Y_X(\tau) - 2 \iu \Im \tau \del_\tau Y_X \bigr) \bigl( (h+ k_\Xi) \ol{Y_\Xi(\tau)} + 2 \iu \Im \tau \ol{\del_\tau Y_\Xi(\tau)} \bigr)- h Y_X(\tau) \ol{Y_\Xi(\tau)}
        \Bigr\}
        \biggr],
\end{align}
where $Y_X$ and $Y_\Xi$ are corresponding to $Y^{(4)}_{\bm{1}}$, $Y^{(4)}_{\bm{1}'}$, or $Y^{(6)}_{\bm{1}}$, depending on the models.
Note that $y$ differs by $c$ from that exhibited in Sec.~\ref{sec:DS-case}.
The mass eigenvalues can be roughly estimated as
\begin{align}
    m_\pm^2 \sim \Lambda^4 
    \times 
    \biggl(
        \frac{|Y_X(\tau)|^2}{\Lambda_X^2} + |Y_X(\tau) + \partial_\tau Y_X(\tau)|^2 
    \biggr) 
    \sim (10 \Lambda^2)^2,
\end{align}
where $Y_X(\tau) \sim c Y_\Xi(\tau)$ and $\Lambda_X \sim \Lambda_\Xi \sim 1$ are assumed.
Then, the masses of $X$-fields can be comparable to that of $\tau$.

\begin{figure}[th]
	\centering
	\includegraphics[width=0.48\textwidth]{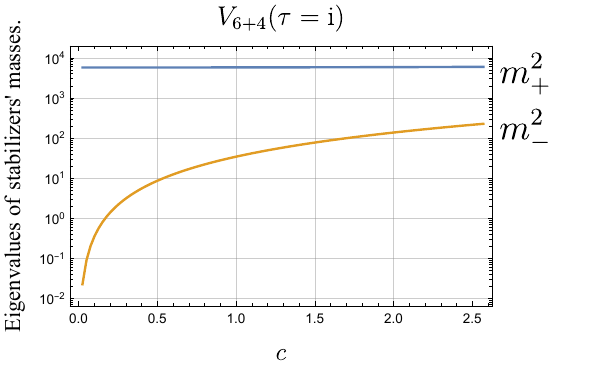}
	\quad 
	\includegraphics[width=0.48\textwidth]{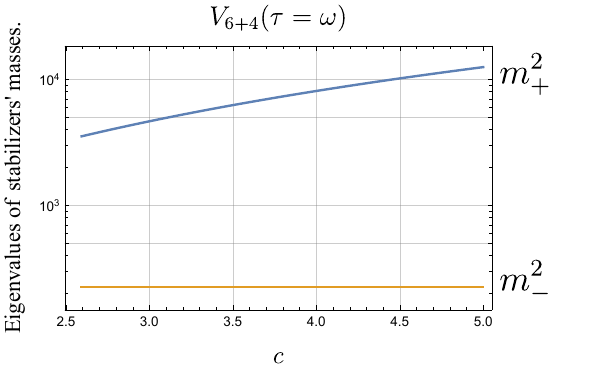}
	\caption{
		(Left) Eigenvalues of stabilizers' masses for $V_{6+4}$ model at $\tau=\iu$ for a small $c$.
		(Right) The same figure at $\tau=\omega$ for $c \gtrsim 2$. Note that $m_-^2$ is independent of $c$ owing to the nature of modular forms at $\tau = \omega$. 
	}
	\label{fig:Y6Y4-StabilizerMasss}
\end{figure}

\begin{figure}[th]
	\centering
	\includegraphics[width=0.48\textwidth]{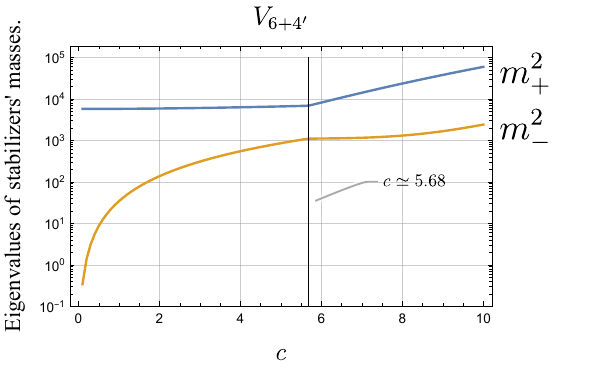}
	\quad 
	\includegraphics[width=0.48\textwidth]{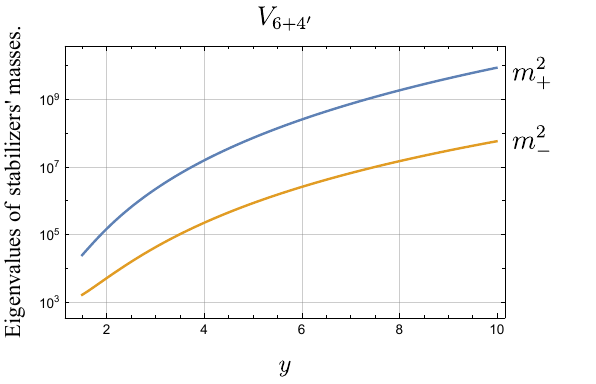}
	\caption{
		(Left) Eigenvalues of stabilizers' masses for $V_{6+4'}$ model at $\tau=\iu$ for $c \lesssim 5.68$, whereas the same figure at $\tau=\iu y$ for $c \gtrsim 5.68$, where $y$ is given by Eq.~\eqref{eq:yofc-V6+4p-taui-2}.
		(Right) The same figure at $\tau=\iu y$, where $y>1$ is approximately given by the vacuum solution of Eq.~\eqref{eq:VY6Y4p-minimum-inf}.
	}
	\label{fig:Y6Y4p-StabilizerMasss}
\end{figure}

\begin{figure}[th]
	\centering
	\includegraphics[width=0.48\textwidth]{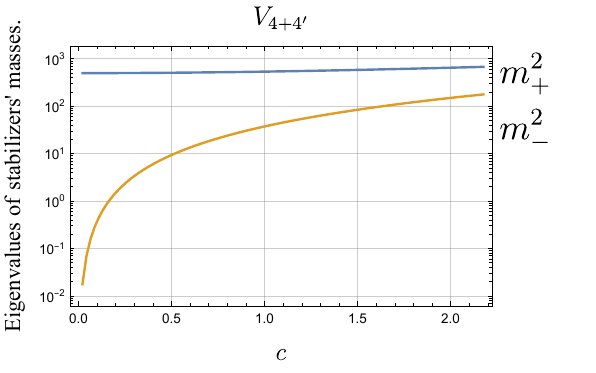}
	\quad 
	\includegraphics[width=0.48\textwidth]{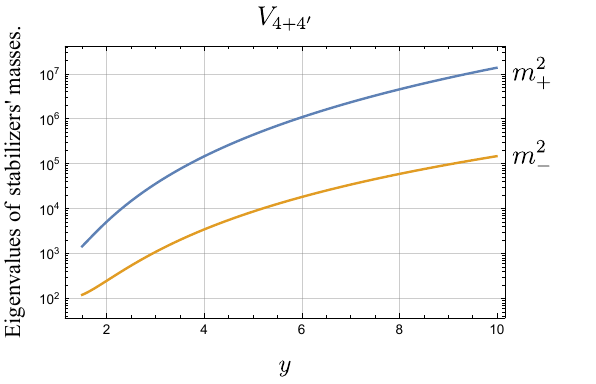}
	\caption{
		(Left) Eigenvalues of stabilizers' masses for $V_{4+4'}$ model at $\tau=\omega$.
		(Right) The same figure at the vacuum of $\tau = -1/2 + \iu y$, where $y$ is approximately given by the solution of Eq.~\eqref{eq:minimum-44p}.
	}
	\label{fig:Y4Y4p-StabilizerMasss}
\end{figure}

In the plots of Figs.~\ref{fig:Y6Y4-StabilizerMasss}, \ref{fig:Y6Y4p-StabilizerMasss}, and \ref{fig:Y4Y4p-StabilizerMasss}, we take $h=2$ and $\Lambda_X = \Lambda_\Xi = 1$ for simplicity.
All the parameters of the physical mass squared turn out to be positive definite.
Hence, the vacua studied in the previous sections are stable.


\section{Conclusion}
\label{sec:conclusion}

In this paper, we study the moduli stabilization in the modular symmetric SUGRA with (approximate) $\rU(1)_{R}$ symmetry, in  which the superpotential consists of multiple modular form coupled to the matter (stabilizer) fields.
First, we study some properties of stabilizer fields and show that the vacuum becomes dS or Minkowski if we have multiple modular forms in the superpotential.
Our models focuses on the cases with one or two modular forms coupled to the stabilizer fields.
For the single stabilizer case, we show that there exists a vacuum at the zero point of the modular form, $\tau = \iu,\ \omega$ and $\iu \infty$, depending on the weight and representation of the modular form. The zero points are also regarded as the fixed point of the modular symmetry in the fundamental domain.
For two stabilizer fields coupled to modular forms, the vacuum can deviate from the fixed points when non-trivial singlet modular form is added to the potential. 
We obtain dS vacua owing to the non-vanishing modular form contribution.
We used perturbations of the modulus against the fixed points to study these shifts of vacua analytically, and confirmed the resultant VEV of modulus also numerically.
As a consequence, there can exist the vacuum at $\Im \tau \gg 1$, and all the vacua obtained in the paper are CP-conserving.

In string theory, there will exist many moduli other than $\tau$, which is focused in this paper.
If the potential scale $\Lambda$ that is associated also with modulus- and $X$-masses is much larger than scales relevant to stabilization terms of such moduli, our stabilization mechanism works well. 
This is because such moduli will be much lighter than $\tau$ and a shift of $\tau$ VEV which can be caused by light moduli will be very small \cite{Abe:2006xi}.

Several issues which we did not discuss in this paper will be left.
A global $\rU(1)_R$ symmetry will be absent in quantum gravity since it is a global symmetry, and is explicitly violated by $W_0(\tau)$ in the superpotential.
In such models, CP violation can be also obtained at the vacuum owing to a competition between $W_0(\tau)$ and $Y'(\tau)X$ in the superpotential \cite{Knapp-Perez:2023nty}, where $Y'(\tau)$ is a modular function.
As another approach to CP violation, there may exist (explicit) violation of the modular symmetry on top of $\rU(1)_R$ violation, which is motivated in an UV completed model. 
In such models, CP violation is caused at the vacuum by an interference between the violation term and $Y(\tau)X$ in the superpotential \cite{Higaki:2024pql}, where $Y(\tau)$ is a modular form.
In this paper we assume the K\"{a}hler potential given by Eq.~\eqref{eq:total-Kahler-potential}, but we can also consider the following additional term to the K\"{a}hler potential \cite{Chen:2019ewa}:
\begin{align}
	\delta \mc{K} = \sum_a \frac{|Y^{(k_a)}_a X_a|^2}{(- \iu \tau + \iu \bar{\tau})^{-h}},
\end{align}
which is introduced to realize a consistent inflationary scenario in Ref.~\cite{Abe:2023ylh}.
This term can have a large contribution in a large $\Im \tau$ region.
A general study of the moduli stabilization including $W_0(\tau)$ and additional corrections can be essential in quantum gravity.
The dS construction motivated by the modular symmetry in string theory attracts a lot of attention and is discussed in Refs.~\cite{Font:1990nt,Ferrara:1990ei,Cvetic:1991qm,Parameswaran:2010ec,Gonzalo:2018guu,Leedom:2022zdm}.
In this work, we show that multiple contributions of the modular forms realize a zero or positive cosmological constant.
The derivation of the superpotential \eqref{eq:total-superpotential} in a string model can be an interesting topic left for future work.

\section*{Acknowledgments}
\noindent
The work of Y.A. is supported by JSPS Overseas Research Fellowships.
This work is supported in part by the Grant-in-Aid for Scientific Research from the Ministry of Education, Science, Sports and Culture (MEXT), Japan No. JP22K03601 (T.H.), JP23K03375 (T.K.) and JP24KJ0249 (K.N.).

\appendix

\section{$A_4$ modular modular forms and their derivatives}

\begin{table}
    \centering
    \begin{tabular}{|c||c|c|c|}\hline
         & $\iu \infty$ & $\iu$ & $\omega$ \\
         \hhline{|=#=|=|=|}
         $Y^{(4)}_{\bm{1}}$ & $1$ & $1.45576$ & $0$ \\
         \hline
         $Y^{(4)}_{\bm{1}'}$ & $0$ & $-1.45576$ & $-1.01247 + 1.75367 \iu$ \\
         \hline
         $Y^{(6)}_{\bm{1}}$ & $1$ & $0$ & $2.88154$
         \\
         \hline
    \end{tabular}
    \caption{
    	The values of the modular forms at fixed points.
    }
    \label{tab:list-modular-forms}
\end{table}

In this appendix, we summarize the features of the modular forms $Y^{(4)}_{\bm{1}}$, $Y^{(4)}_{\bm{1}'}$, and $Y^{(6)}_{\bm{1}}$ that we use in the main part of this paper.
The modular forms values at the fixed points $\tau = \iu,\ \omega,\ \iu \infty$ are listed in Tab.~\ref{tab:list-modular-forms}.

\subsection{Modular form around $\tau = \iu$}
\label{sec:result-taui-expansion}

As discussed in Sec.~\ref{sec:expansion-tau-iu}, the $l$-th derivative of the modular form with the weight $k$ satisfies 
\begin{align}
	(\iu^{2l} - \iu^{-k}) \frac{\dd^l \tilde{Y}^{(k)}_{\bm{r}}}{\dd s^l} \bigg|_{s \to 0 } =0,
\end{align}
where $\tilde{Y}^{(k)}_{\bm{r}}(s) = (1-s)^k Y^{(k)}_{\bm{r}}(\tau)$ and $s =(\tau - \iu)/(\tau + \iu)$.
The modular form is expanded as 
\begin{align}
	Y^{(k)}_{\bm{r}} = \sum_{l=0} \frac{1}{l!} \frac{\dd^l Y^{(k)}_{\bm{r}}}{\dd \tau^l}(\iu) (\tau - \iu)^l,
\end{align}
and the derivatives are related to $\tilde{Y}^{(k)}_{\bm{r}}(s)$ and its derivatives as 
\begin{align}
	\frac{\dd Y^{(k)}_{\bm{r}}}{\dd \tau} &= \frac{\iu (1-s)^{1+k}}{2}\biggl[
		k \tilde{Y}^{(k)}_{\bm{r}} - (1-s) \frac{\dd \tilde{Y}^{(k)}_{\bm{r}}}{\dd s}
	\biggr],
	\\
	\frac{\dd^2 Y^{(k)}_{\bm{r}}}{\dd \tau^2} &= - \frac{(1-s)^{2+k}}{4} \biggl[
		k(1+k)\tilde{Y}^{(k)}_{\bm{r}} - (1-s) \Bigl(
			2 (1+k) \frac{\dd \tilde{Y}^{(k)}_{\bm{r}}}{\dd s} - ( 1- s) \frac{\dd^2 \tilde{Y}^{(k)}_{\bm{r}}}{\dd s^2}
		\Bigr)
	\biggr],
	\\
	\frac{\dd^3 \tilde{Y}^{(k)}_{\bm{r}}}{\dd \tau^3} &=- \frac{\iu (1-s)^{3+k}}{8} \biggl[
		k(2+3k +k^2) \tilde{Y}^{(k)}_{\bm{r}} - (1-s) \Bigl\{
			3 (2 + 3k + k^2) \frac{\dd \tilde{Y}^{(k)}_{\bm{r}}}{\dd s} 
			\nn \\
			& \qquad 
			- (1-s) \Bigl(
				3 (2 +k) \frac{\dd^2 \tilde{Y}^{(k)}_{\bm{r}}}{\dd s^2} - (1-s) \frac{\dd^3 \tilde{Y}^{(k)}_{\bm{r}}}{\dd s^3} 
			\Bigr)
		\Bigr\}
	\biggr],
	\\
	\frac{\dd^4 Y^{(k)}_{\bm{r}}}{\dd \tau^4} &= \frac{(1-s)^{4+k}}{16} \biggl[
		k(6 + 11k + 6k^2 + k^3) \tilde{Y}^{(k)}_{\bm{r}} - (1-s) \Bigl\{
			4 (6 + 11k + 6k^2 + k^3) \frac{\dd \tilde{Y}^{(k)}_{\bm{r}}}{\dd s} 
			\nn \\
			& \qquad - (1-s) \Bigl(
				6(6 + 5k + k^2) \frac{\dd^2 \tilde{Y}^{(k)}_{\bm{r}}}{\dd s^2} - (1-s) \Bigl(
					4 ( 3 + k)\frac{\dd^3 \tilde{Y}^{(k)}_{\bm{r}}}{\dd s^3} - (1-s) \frac{\dd^4 \tilde{Y}^{(k)}_{\bm{r}}}{\dd s^4}
				\Bigr)
			\Bigr)
		\Bigr\}
	\biggr].
\end{align}
In the left-hand side, $\tilde{Y}^{(k)}_{\bm{r}}$ is a function of $s$.
We note that CP transformation acts on the modular form as \cite{Novichkov:2019sqv}
\begin{align}
	\mr{CP}: Y^{(k)}_{\bm{1}^{(\prime)}}(\tau) \mapsto Y^{(k)}_{\bm{1}^{(\prime)}}(- \bar{\tau}) = \ol{Y^{(k)}_{\bm{1}^{(\prime)}}}(\tau),
\end{align}
which implies that the modular forms take real values at $\tau = \iu$.
From these equations, we obtain the following expansions of the weight $4$ and $6$ modular forms around the fix point $\tau = \iu$.

\paragraph{Weight 4.}

\begin{align}
	Y^{(4)}_{\bm{r}} &= 
	\tilde{Y}^{(4)}_{\bm{r}}(0) 
	+ 2 \iu \tilde{Y}^{(4)}_{\bm{r}}(0) (\tau - \iu)
	+ \frac{1}{2} \biggl[
		- 5 \tilde{Y}^{(4)}_{\bm{r}} - \frac{1}{4} \frac{\dd^2 \tilde{Y}^{(4)}_{\bm{r}}}{\dd s^2}(0)
	\biggr] (\tau - \iu)^2
	\nn \\
	& \quad 
	- \frac{\iu}{3!} \biggl[
		15 \tilde{Y}^{(4)}_{\bm{r}}(0) + \frac{9}{4} \frac{\dd^2 \tilde{Y}^{(4)}_{\bm{r}}}{\dd s^2}(0)
	\biggr] (\tau - \iu)^3
	\nn \\
	& \quad 
	+ \frac{1}{4!} \biggl[
		\frac{105}{2} \tilde{Y}^{(4)}_{\bm{r}}(0) + \frac{63}{4} \frac{\dd^2 \tilde{Y}^{(4)}_{\bm{r}}}{\dd s^2}(0) + \frac{1}{16} \frac{\dd^4 \tilde{Y}^{(4)}_{\bm{r}}}{\dd s^4}(0)
	\biggr] (\tau - \iu)^4
	+ \mc{O}((\tau- \iu)^5) ,
\end{align}
with $\bm{r} = \bm{1},\ \bm{1}'$.

\paragraph{Weight 6.}

\begin{align}
	Y^{(6)}_{\bm{1}} &= 
	- \frac{\iu}{2} \frac{\dd \tilde{Y}^{(6)}_{\bm{1}}}{\dd s}(0) (\tau - \iu)
	+ \frac{7}{4} \frac{\dd \tilde{Y}^{(6)}_{\bm{r}}}{\dd s}(0) (\tau - \iu)^2
	\nn \\
	& \quad 
	+ \frac{\iu}{3!} \biggl[
		21 \frac{\dd \tilde{Y}^{(6)}_{\bm{1}}}{\dd s}(0) + \frac{1}{8} \frac{\dd^3 \tilde{Y}^{(6)}_{\bm{1}}}{\dd s^3}(0)
	\biggr] (\tau- \iu )^3
	\nn \\
	& \quad 
	+ \frac{1}{4!} \biggl[
		- 126 \frac{\dd \tilde{Y}^{(6)}_{\bm{1}}}{\dd s}(0) - \frac{9}{4} \frac{\dd^3 \tilde{Y}^{(6)}_{\bm{1}}}{\dd s^3}(0)
	\biggr] (\tau - \iu )^4 
	+ \mc{O}((\tau-\iu)^5).
\end{align}

\subsection{Modular form around $\tau = \omega$}
\label{sec:result-tauomega-expansion}

As shown in Sec.~\ref{sec:expansion-tau-omega}, the $l$-th derivative of the modular form is characterized by 
\begin{align}
	(\omega^{2l} - \omega^{q_{\bm{r}} - k} ) \frac{\dd^l \hat{Y}^{(k)}_{\bm{r}}}{\dd u^l} =0,
\end{align}
where we introduce $u = (\tau-\omega)/(\tau - \omega^2)$ and $Y^{(k)}_{\bm{r}} = (1-u)^k \hat{Y}^{(k)}_{\bm{r}}$.
$q_{\bm{r}}$ takes $0$ and $1$ for the singlet $\bm{1}$ and the non-trivial singlet $\bm{1}'$, respectively.
The modular form is expanded as 
\begin{align}
	Y^{(k)}_{\bm{r}} = \sum_{l=0} \frac{1}{l!} \frac{\dd^l Y^{(k)}_{\bm{r}}}{\dd \tau^l}(\omega ) (\tau - \omega)^l,
\end{align}
and the derivatives are given by
\begin{align}
	\frac{\dd Y^{(k)}_{\bm{r}}}{\dd \tau} &= \frac{\iu (1-u)^{1+k}}{\sqrt{3}}\biggl[
		k \hat{Y}^{(k)}_{\bm{r}} - (1-u) \frac{\dd \hat{Y}^{(k)}_{\bm{r}}}{\dd u}
	\biggr],
	\\
	\frac{\dd^2 Y^{(k)}_{\bm{r}}}{\dd \tau^2}&= - \frac{(1-u)^{2+k}}{3} \biggl[
		k(1+k) \hat{Y}^{(k)}_{\bm{r}}- (1-u) \Bigl(
			2 (1 + k) \frac{\dd \hat{Y}^{(k)}_{\bm{r}}}{\dd u} - (1-u) \frac{\dd^2 \hat{Y}^{(k)}_{\bm{r}}}{\dd u^2}
		\Bigr)
	\biggr],
	\\
	\frac{\dd^3 Y^{(k)}_{\bm{r}}}{\dd \tau^3} &= - \frac{\iu (1-u)^{3+k}}{3 \sqrt{3}} \biggl[
		k ( 2 +3k +k^2) \hat{Y}^{(k)}_{\bm{r}} - (1-u) \Bigl\{
			3 (2+3k+k^2) \frac{\dd \hat{Y}^{(k)}_{\bm{r}}}{\dd u}
			\nn \\
			& \qquad - ( 1-u) \Bigl(
				3 (2+k) \frac{\dd^2 \hat{Y}^{(k)}_{\bm{r}}}{\dd u^2} - (1-u) \frac{\dd^3 \hat{Y}^{(k)}_{\bm{r}}}{\dd u^3}
			\Bigr)
		\Bigr\}
	\biggr]
	\\
	\frac{\dd^4 Y^{(k)}_{\bm{r}}}{\dd \tau^4} &= \frac{(1-u)^{4+k}}{9} \biggl[
		k(6 + 11 k + 6k^2 + k^3)\hat{Y}^{(k)}_{\bm{r}} - (1-u) \Bigl\{
			4 (6 + 11 k + 6k^2 + k^3) \frac{\dd \hat{Y}^{(k)}_{\bm{r}}}{\dd u} 
			\nn \\
			& \qquad 
			- (1-u) \Bigl(
				6 ( 6 + 5k + k^2) \frac{\dd^2 \hat{Y}^{(k)}_{\bm{r}}}{\dd u^2} - ( 1-u) \Bigl(
					4(3+k) \frac{\dd^3 \hat{Y}^{(k)}_{\bm{r}}}{\dd u^3} - (1-u) \frac{\dd^4 \hat{Y}^{(k)}_{\bm{r}}}{\dd u^4}
				\Bigr)
			\Bigr)
		\Bigr\}
	\biggr].
\end{align}
$\hat{Y}^{(k)}_{\bm{r}}$ in the left-hand side is a function of $u$.
From these we obtain the following expansions of the modular forms.

\paragraph{Weight 4 trivial singlet.}

\begin{align}
	Y^{(4)}_{\bm{1}} &= 
	- \frac{\iu}{\sqrt{3}} \frac{\dd \hat{Y}^{(4)}_{\bm{1}}}{\dd u}(0) (\tau - \omega)
	+ \frac{5}{3} \frac{\dd \hat{Y}^{(4)}_{\bm{1}}}{\dd u}(0) (\tau - \omega)^2
	\nn \\
	& \quad 
	+\frac{\iu}{3!} \biggl[
		\frac{30}{\sqrt{3}} \frac{\dd \hat{Y}^{(4)}_{\bm{1}}}{\dd u}(0)
	\biggr](\tau - \omega)^3
	\nn \\
	& \quad 
	+ \frac{1}{4!} \biggl[
		- \frac{280}{3} \frac{\dd \hat{Y}^{(4)}_{\bm{1}}}{\dd u}(0) + \frac{1}{9} \frac{\dd^4 \hat{Y}^{(4)}_{\bm{1}}}{\dd u^4}(0)
	\biggr] (\tau - \omega)^4 
	+ \mc{O}((\tau- \omega)^5).
\end{align}

\paragraph{Weight 4 non-trivial singlet.}

\begin{align}
	Y^{(4)}_{\bm{1}'} &= 
	\hat{Y}^{(4)}_{\bm{1}'}(0)
	+ \frac{4 \iu}{\sqrt{3}} \hat{Y}^{(4)}_{\bm{1}'}(0)  (\tau - \omega)
	- \frac{10}{3} \hat{Y}^{(4)}_{\bm{1}'}(0)(\tau - \omega)^2
	\nn \\
	& \quad 
	+\frac{\iu}{3!} \biggl[
		- \frac{40}{\sqrt{3}} \hat{Y}^{(4)}_{\bm{1}'}(0) + \frac{1}{3 \sqrt{3}} \frac{\dd^3 \hat{Y}^{(4)}_{\bm{1}'}}{\dd u^3} (0)
	\biggr] (\tau - \omega)^3
	\nn \\
	& \quad 
	+ \frac{1}{4!} \biggl[
		\frac{280}{3} \hat{Y}^{(4)}_{\bm{1}'}(0) - \frac{28}{9} \frac{\dd^3 \hat{Y}^{(4)}_{\bm{1}'}}{\dd u^3}(0)
	\biggr] (\tau - \omega)^4
	+ \mc{O}((\tau - \omega)^5).
\end{align}

\paragraph{Weight 6 trivial singlet.}

\begin{align}
	Y^{(6)}_{\bm{1}} &= 
	\hat{Y}^{(6)}_{\bm{1}}(0)
	+ 2\sqrt{3} \iu \hat{Y}^{(6)}_{\bm{1}}(0) (\tau - \omega)
	- 7 \hat{Y}^{(6)}_{\bm{1}}(\tau - \omega)^2
	\nn \\
	& \quad 
	+ \frac{\iu}{3!} \biggl[
		- \frac{112}{\sqrt{3}} \hat{Y}^{(6)}_{\bm{1}}(0) + \frac{1}{3 \sqrt{3}} \frac{\dd^3 \hat{Y}^{(6)}_{\bm{1}}}{\dd u^3}(0)
	\biggr] (\tau - \omega)^3
	\nn \\
	& \quad 
	+ \frac{1}{4!}\biggl[
		336 \hat{Y}^{(6)}_{\bm{1}}(0) - 4 \frac{\dd^3 \hat{Y}^{(6)}_{\bm{1}}}{\dd u^3}(0)
	\biggr] (\tau - \omega)^4
	+ \mc{O}((\tau-\omega)^5).
\end{align}

\subsection{$q$ expansion of modular form and its derivative}
\label{sec:result-q-expansion}

In the $\Im \tau \gg 1$ region, the modular forms are expanded by $q = \exp ( 2\pi \iu \tau)$~\cite{Okada:2020ukr,Feruglio:2022koo,Kikuchi:2023uqo,Novichkov:2021evw} as 
\begin{align}
    Y^{(4)}_{\bm{1}} (\tau) &= 1 + 240 q + 2160 q^2 + 6720 q^3 + \mc{O}(q^4),
    \\
    Y^{(4)}_{\bm{1}'}(\tau) &= q^{1/3} ( -12 q + 96 q^2  - 230 q^3 + \mc{O}(q^4) ),
    \\
    Y^{(6)}_{\bm{1}} (\tau) &= 1 - 504 q - 16632 q^2 - 122976 q^3 + \mc{O}(q^4),
\end{align}
which is called the $q$-expansion.
The derivatives are systematically given by these formulae and listed below.

\paragraph{Weight 4 trivial singlet.}

\begin{align}
    \frac{\dd Y^{(4)}_{\bm{1}}}{ \dd \tau} &= 2 \pi \iu ( 240 q + 4320 q^2 + 20160 q^3 + \mc{O}(q^4) ),
    \\
    \frac{\dd^2 Y^{(4)}_{\bm{1}}}{\dd \tau^2} &= - 960 \pi^2 q - 34560 \pi^2 q^2 -40320 \pi^i q^3 + \mc{O}(q^4),
    \\
    \frac{\dd^3 Y^{(4)}_{\bm{1}}}{\dd \tau^3} &= -1920\iu \pi^3 q - 138240 \iu \pi^3 q^2 - 1451520 \iu \pi^3 q^3 + \mc{O}(q^4),
    \\
    \frac{\dd^4 Y^{(4)}_{\bm{1}}}{\dd \tau^4} &= 3840 \pi^4 q + 552960 \pi^4 q^2 + 8709120 \pi^4 q^3 + \mc{O}(q^4).
\end{align}

\paragraph{Weight 4 non-trivial singlet.}

\begin{align}
    \frac{\dd Y^{(4)}_{\bm{1}'}}{\dd \tau} &= - 2\pi \iu q^{1/3} ( 4 + 128 q + 560 q^{2} + \mc{O}(q^3) ),
    \\
    \frac{\dd^2 Y^{(4)}_{\bm{1}'}}{\dd \tau^2} &= \frac{\pi^2 q^{1/3}}{3} (16 - 2048 q + 15680 q^2 + \mc{O}(q^3)),
    \\
    \frac{\dd^3 Y^{(4)}_{\bm{1}'}}{\dd \tau^3} &= \frac{\iu \pi^3 q^{1/3}}{9} (32 - 16384 q + 219520 q^2 + \mc{O}(q^3)),
    \\
    \frac{\dd^4 Y^{(4)}_{\bm{1}'}}{\dd \tau^4} &=- \frac{\pi^4 q^{1/3}}{27} (-64 + 131072 q - 3073280 q^2 + \mc{O}(q^3)).
\end{align}

\paragraph{Wight 6 trivial singlet.}

\begin{align}
    \frac{\dd Y^{(6)}_{\bm{1}}}{\dd \tau} &= - 2\pi \iu ( 504 q + 332604 q^2 + 368928 q^3 + \mc{O}(q^4) ),
    \\
    \frac{\dd^2 Y^{(6)}_{\bm{1}}}{\dd \tau^2} &= 2016 \pi^2 q + 266112 \pi^2 q^2 + 4427136 \pi^2 q^3 + \mc{O}(q^4),
    \\
    \frac{\dd^3 Y^{(6)}_{\bm{1}}}{\dd \tau^3} &= 4032 \pi^3 \iu q + 106448 \pi^3 \iu q^2 + 26562816 \pi^3 \iu q^3 + \mc{O}(q^4),
    \\
    \frac{\dd^4 Y^{(6)}_{\bm{1}}}{\dd \tau^4} &= - 8064 \pi^4 q - 4257792 \pi^4 q^2 - 159376896 \pi^4 q^3 + \mc{O}(q^4).
\end{align}

\section{Details of potential coefficients}
\label{sec:details-of-potential}

In this appendix, we discuss the perturbative expansion of a single modular form potential:
\begin{align}
	V^{(k)}_{\bm{r}} =(2 \Im \tau)^{k} |Y^{(k)}_{\bm{r}}|^2,
\end{align}
where $k$ is the modular weight of the modular form $Y^{(k)}_{\bm{r}}$ and $\bm{r}$ denotes the representation.
We show the fourth order perturbation around the fix points $\tau = \iu,\ \omega$.

\subsection{Potential expansion around $\tau = \iu$}

\subsubsection{Potential coefficients}

We introduce the perturbation coefficient by
\begin{align}
	V^{(k)}_{\bm{r}}= \sum_{i,j} C^{(k)}_{\bm{r}}[i:j] x^i y^j,
	\label{eq:Vkrexpansion-aroundi}
\end{align}
where the modulus is expanded as $\tau = \iu + x + \iu y$.

\paragraph{Weight 4 trivial and non-trivial singlets.}

For the weight 4 singlets, the coefficients \eqref{eq:Vkrexpansion-aroundi} are given by 
\begin{subequations}
\begin{align}
	C^{(4)}_{\bm{r}}[0:0] &=16 |\tilde{Y}^{(4)}_{\bm{r}}|^2,
	\\
	C^{(4)}_{\bm{r}}[0:2] &= 
		4\frac{\dd^2 \tilde{Y}^{(4)}_{\bm{r}}}{\dd s^2} \tilde{Y}^{(4)}_{\bm{r}}  - 16 |\tilde{Y}^{(4)}_{\bm{r}}|^2,
	\\
	C^{(4)}_{\bm{r}}[0:3] &=
		-4 \frac{\dd^2 \tilde{Y}^{(4)}_{\bm{r}}}{\dd s^2} \tilde{Y}^{(4)}_{\bm{r}}  + 16 |\tilde{Y}^{(4)}_{\bm{r}}|^2,
	\\
	C^{(4)}_{\bm{r}}[0:4] &=
		- \frac{\dd^2 \tilde{Y}^{(4)}_{\bm{r}}}{\dd s^2} \tilde{Y}^{(4)}_{\bm{r}}
        + \frac{1}{12} \frac{\dd^4 \tilde{Y}^{(4)}_{\bm{r}}}{\dd s^4} \tilde{Y}^{(4)}_{\bm{r}}
        + \frac{1}{4} \left| \frac{\dd^2 \tilde{Y}^{(4)}_{\bm{r}}}{\dd s^2} \right|^2
        - 6 |\tilde{Y}^{(4)}_{\bm{r}}|^2,
	\\
	C^{(4)}_{\bm{r}}[1:1] &=4 \iu \biggl[
		\ol{\frac{\dd^2 \tilde{Y}^{(4)}_{\bm{r}}}{\dd s^2}} \tilde{Y}^{(4)}_{\bm{r}} - \frac{\dd^2 \tilde{Y}^{(4)}_{\bm{r}}}{\dd s^2} \ol{\tilde{Y}^{(4)}_{\bm{r}}}
	\biggr] = 0,
	\\
	C^{(4)}_{\bm{r}}[1:2] &= 6 \iu \biggl[
		- \ol{\frac{\dd^2 \tilde{Y}^{(4)}_{\bm{r}}}{\dd s^2}} \tilde{Y}^{(4)}_{\bm{r}} + \frac{\dd^2 \tilde{Y}^{(4)}_{\bm{r}}}{\dd s^2} \ol{\tilde{Y}^{(4)}_{\bm{r}}}
	\biggr] = 0,
	\\
	C^{(4)}_{\bm{r}}[1:3] &= 
        \frac{\iu}{6} \biggl[
		12 \ol{\frac{\dd^2 \tilde{Y}^{(4)}_{\bm{r}}}{\dd s^2}} \tilde{Y}^{(4)}_{\bm{r}} 
        - 12 \frac{\dd^2 \tilde{Y}^{(4)}_{\bm{r}}}{\dd s^2} \ol{\tilde{Y}^{(4)}_{\bm{r}}} 
        + \ol{\frac{\dd^4 \tilde{Y}^{(4)}_{\bm{r}}}{\dd s^4}} \tilde{Y}^{(4)}_{\bm{r}} 
        - \frac{\dd^4 \tilde{Y}^{(4)}_{\bm{r}}}{\dd s^4} \ol{\tilde{Y}^{(4)}_{\bm{r}}}
	\biggr] = 0,
	\\
	C^{(4)}_{\bm{r}}[2:0] &= 
        -4 \frac{\dd^2 \tilde{Y}^{(4)}_{\bm{r}}}{\dd s^2} \tilde{Y}^{(4)}_{\bm{r}}
        + 16 |\tilde{Y}^{(4)}_{\bm{r}}|^2,
	\\
	C^{(4)}_{\bm{r}}[2:1] &=
		12 \frac{\dd^2 \tilde{Y}^{(4)}_{\bm{r}}}{\dd s^2} \tilde{Y}^{(4)}_{\bm{r}} + 16 |\tilde{Y}^{(4)}_{\bm{r}}|^2,
	\\
	C^{(4)}_{\bm{r}}[2:2] &= \frac{1}{2} \left| \frac{\dd^2 \tilde{Y}^{(4)}_{\bm{r}}}{\dd s^2} \right|^2 - 18 \frac{\dd^2 \tilde{Y}^{(4)}_{\bm{r}}}{\dd s^2} \tilde{Y}^{(4)}_{\bm{r}} - \frac{1}{2} \frac{\dd^4 \tilde{Y}^{(4)}_{\bm{r}}}{\dd s^4}  \tilde{Y}^{(4)}_{\bm{r}} + 4 |\tilde{Y}^{(4)}_{\bm{r}}|^2
	\\
	C^{(4)}_{\bm{r}}[3:0] &= 2 \iu \biggl[
		\ol{\frac{\dd^2 \tilde{Y}^{(4)}_{\bm{r}}}{\dd s^2}} \tilde{Y}^{(4)}_{\bm{r}} - \frac{\dd^2 \tilde{Y}^{(4)}_{\bm{r}}}{\dd s^2} \ol{\tilde{Y}^{(4)}_{\bm{r}}}
	\biggr] = 0
	\\
	C^{(4)}_{\bm{r}}[3:1] &= \frac{\iu}{6} \biggl[
		- 60 \ol{\frac{\dd^2 \tilde{Y}^{(4)}_{\bm{r}}}{\dd s^2}} \tilde{Y}^{(4)}_{\bm{r}} - \ol{\frac{\dd^4 \tilde{Y}^{(4)}_{\bm{r}}}{\dd s^4}} \tilde{Y}^{(4)}_{\bm{r}}
		+ 60 \frac{\dd^2 \tilde{Y}^{(4)}_{\bm{r}}}{\dd s^2} \ol{\tilde{Y}^{(4)}_{\bm{r}}} + \frac{\dd^4 \tilde{Y}^{(4)}_{\bm{r}}}{\dd s^4} \ol{\tilde{Y}^{(4)}_{\bm{r}}}
	\biggr] = 0
	\\
	C^{(4)}_{\bm{r}}[4:0] &= 
		7 \frac{\dd^2 \tilde{Y}^{(4)}_{\bm{r}}}{\dd s^2} \tilde{Y}^{(4)}_{\bm{r}}
        + \frac{1}{12} \frac{\dd^4 \tilde{Y}^{(4)}_{\bm{r}}}{\dd s^4} \tilde{Y}^{(4)}_{\bm{r}}
        + \frac{1}{4} \left| \frac{\dd^2 \tilde{Y}^{(4)}_{\bm{r}}}{\dd s^2} \right|^2 
		+ 10 |\tilde{Y}^{(4)}_{\bm{r}}|^2
	\\
	C^{(4)}_{\bm{r}}[0:1]&= C^{(4)}_{\bm{r}}[1:0]=0,
\end{align}
\label{eq:coefficients-V4r-taui}
\end{subequations}
where $\bm{r} = \bm{1},\ \bm{1}'$.
The numerical values of these coefficients are summarized in Tabs.~\ref{tab:table-C4bm1} and \ref{tab:table-C4bm1p}.
The higher order fluctuation than the fourth-order are noted as -- in these tables and the same notation is used in the following parts.
The explicit forms of the potential expansion become follows:
\begin{align}
	V^{(4)}_{\bm{1}}(\iu + x + \iu y)
        &\approx 33.9
            + 237\, y^2 
            - 237\, y^3
            + 622\, y^4 
            \nn \\
            &\quad
            + \left(
                - 305 
                + 846\,y 
                - 1.10\times10^3\,y^2 \right) \, x^2
            + 1.20\times10^3 \, x^4,
            \label{eq:V4_1_expansion_4thorder_i}
    \\
    V^{(4)}_{\bm{1}'}(\iu + x + \iu y)
        &\approx 33.9
            - 88.0\, y^2 
            + 88.0\, y^3
            + 11.6\, y^4 
            \nn \\
            &\quad
            + \left(
                20.2 
                + 128\,y 
                + 360\,y^2 \right) \, x^2
            - 62.7 \, x^4.
            \label{eq:V4_1p_expansion_4thorder_i}
\end{align}
\begin{table}[t]
    \centering
    \begin{tabular}{|c||c|c|c|c|c|}
        \hline
         $i \backslash j$ & $0$ & $1$ & $2$ & $3$ & $4$
         \\
         \hhline{|=#=|=|=|=|=|}
         $0$ & $33.908$ & $0$ & $236.75$ & $-236.75$ & $621.76$ 
         \\ \hline
         $1$ & $0$ & $0$ & $0$ & $0$ & --
         \\ \hline
         $2$ & $-304.57$ & $845.88$ & $-1101.5$ & -- & --
         \\ \hline
         $3$ & $0$ & $0$ & -- & -- & --
         \\ \hline
         $4$ & $1197.0$ & -- & -- & -- & --
         \\
         \hline
    \end{tabular}
    \caption{Numerical values of $C^{(4)}_{\bm{1}}[i:j]$.}
    \label{tab:table-C4bm1}
\end{table}
\begin{table}[t]
    \centering
    \begin{tabular}{|c||c|c|c|c|c|}
        \hline
         $i \backslash j$ & $0$ & $1$ & $2$ & $3$ & $4$
         \\
         \hhline{|=#=|=|=|=|=|}
         $0$ & $33.908$ & $0$ & $-88.039$ & $88.039$ & $11.619$ 
         \\ \hline
         $1$ & $0$ & $0$ & $0$ & $0$ & --
         \\ \hline
         $2$ & $20.224$ & $128.49$ & $360.09$ & -- & --
         \\ \hline
         $3$ & $0$ & $0$ & -- & -- & -- 
         \\ \hline
         $4$ & $-62.735$ & -- & -- & -- & --
         \\
         \hline
    \end{tabular}
    \caption{Numerical values of $C^{(4)}_{\bm{1}'}[i:j]$.}
    \label{tab:table-C4bm1p}
\end{table}

\paragraph{Weight 6 singlet.}

For the weight 6 singlet modular form, the coefficients are
\begin{subequations}
\begin{align}
	C^{(6)}_{\bm{1}}[0:2] &= 16 \left| \frac{\dd \tilde{Y}^{(6)}_{\bm{1}}}{\dd s} \right|^2,
	\\
	C^{(6)}_{\bm{1}}[0:3] &=-16 \left| \frac{\dd \tilde{Y}^{(6)}_{\bm{1}}}{\dd s} \right|^2,
	\\
	C^{(6)}_{\bm{1}}[0:4] &= 
		\frac{4}{3} \frac{\dd^3 \tilde{Y}^{(6)}_{\bm{1}}}{\dd s^3} \frac{\dd \tilde{Y}^{(6)}_{\bm{1}}}{\dd s} 
        - 12 \left| \frac{\dd \tilde{Y}^{(6)}_{\bm{1}}}{\dd s} \right|^2,
	\\
	C^{(6)}_{\bm{1}}[1:3] &= 
        \frac{4\iu}{3} \biggl[
		\ol{\frac{\dd^3 \tilde{Y}^{(6)}_{\bm{1}}}{\dd s^3}} \frac{\dd \tilde{Y}^{(6)}_{\bm{1}}}{\dd s} - \frac{\dd^3 \tilde{Y}^{(6)}_{\bm{1}}}{\dd s^3} \ol{\frac{\dd \tilde{Y}^{(6)}_{\bm{1}}}{\dd s}}
	\biggr] = 0,
	\\
	C^{(6)}_{\bm{1}}[2:0]&= 16 \left| \frac{\dd \tilde{Y}^{(6)}_{\bm{1}}}{\dd s} \right|^2,
	\\
	C^{(6)}_{\bm{1}}[2:1]&= - 16 \left| \frac{\dd \tilde{Y}^{(6)}_{\bm{1}}}{\dd s} \right|^2,
	\\
	C^{(6)}_{\bm{1}}[2:2]&= -40 \left| \frac{\dd \tilde{Y}^{(6)}_{\bm{1}}}{\dd s} \right|^2,
	\\
	C^{(6)}_{\bm{1}}[3:1]&= \frac{4\iu}{3} \biggl[
		\ol{\frac{\dd^3 \tilde{Y}^{(6)}_{\bm{1}}}{\dd s^3}} \frac{\dd \tilde{Y}^{(6)}_{\bm{1}}}{\dd s} - \frac{\dd^3 \tilde{Y}^{(6)}_{\bm{1}}}{\dd s^3} \ol{\frac{\dd \tilde{Y}^{(6)}_{\bm{1}}}{\dd s}}
	\biggr] = 0,
	\\
	C^{(6)}_{\bm{1}}[4:0] &= 
		- \frac{4}{3} \frac{\dd^3 \tilde{Y}^{(6)}_{\bm{1}}}{\dd s^3} \frac{\dd \tilde{Y}^{(6)}_{\bm{1}}}{\dd s} 
        - 28 \left| \frac{\dd \tilde{Y}^{(6)}_{\bm{1}}}{\dd s} \right|^2,
	\\
	C^{(6)}_{\bm{1}}[0:0] &= C^{(6)}_{\bm{1}}[0:1] = C^{(6)}_{\bm{1}}[1:0] = C^{(6)}_{\bm{1}}[1:1] = C^{(6)}_{\bm{1}}[1:2] = C^{(6)}_{\bm{1}}[3:0] = 0
\end{align}
\label{eq:coefficients-V61-taui}
\end{subequations}
and their numerical values are listed in Tab.~\ref{tab:table-C6bm1}.
From these, the explicit expansion form of $V^{(6)}_{\bm{1}}$ is given by 
\begin{align}
	V^{(6)}_{\bm{1}}(\iu + x + \iu y)
        \times 10^{-3}
        \approx&~ 2.84 y^2 - 2.84 y^3 + 8.44 y^4     
        \nn \\
            & +\left(
                2.84-2.84 y-7.09 y^2
            \right)x^2
            -15.5 x^4.
            \label{eq:V6expansion_4thorder_i}
\end{align}
\begin{table}[t]
    \centering
    \begin{tabular}{|c||c|c|c|c|c|}\hline
         $i \backslash j$ & $0$ & $1$ & $2$ & $3$ & $4$  
         \\
         \hhline{|=#=|=|=|=|=|}
         $0$ & $0$ & $0$ & $2836.9$ & $-2836.9$ & $8439.7$ 
         \\ \hline
         $1$ & $0$ & $0$ & $0$ & $0$ & --
         \\ \hline
         $2$ & $2836.9$ & $-2836.9$ & $-7092.2$ & -- & --
         \\ \hline
         $3$ & $0$ & $0$ & -- & -- & --
         \\ \hline
         $4$ & $-15532$ & -- & -- & -- & --
         \\
         \hline
    \end{tabular}
    \caption{Numerical values of $C^{(6)}_{\bm{1}}[i:j]$.}
    \label{tab:table-C6bm1}
\end{table}

\subsubsection{Coefficients of potential derivative}

In Sec.~\ref{sec:V-6+4}, we introduced the following notation for the potential derivatives around $\tau = \iu$:
\begin{align}
	\frac{\del V_r}{\del x} &= 4 \mc{X}_4(c) \biggl[ x^2 + \frac{1}{2} \frac{\mc{X}_2(c,y)}{\mc{X}_4(c)} \biggr] x,
	\\
	\frac{\del V_r}{\del y} &= \mc{Y}_1(c,x) + \mc{Y}_2(c,x) y + \mc{Y}_3 (c) y^3 + \mc{Y}_4(c) y^3,
\end{align}
where the label runs $r = 6+4,\ 6 +4'$.
These have following forms with the coefficients \eqref{eq:coefficients-V4r-taui} and \eqref{eq:coefficients-V61-taui}
\begin{align}
    \mc{X}_2(c,y)&= C^{(6)}_{\bm{1}}[2:0] + c^2 \bigl(
        C^{(4)}_{\bm{r}}[2:0] + C^{(4)}_{\bm{r}}[2:1] y + C^{(4)}_{\bm{r}}[2:2] y^2
    \bigr),
    \label{eq:mcX2-app}
    \\
    \mc{X}_4(c) &= C^{(6)}_{\bm{1}}[4:0] + c^2 C^{(4)}_{\bm{r}}[4:0],
    \label{eq:mcX4-app}
    \\
    \mc{Y}_1(c,x) &= \bigl(
        C^{(6)}_{\bm{1}}[2:1]+ c^2 C^{(4)}_{\bm{r}}[2:1]
    \bigr) x^2,
    \label{eq:mcY1-app}
    \\
    \mc{Y}_2(c,x) &= 2 \Bigl[
        C^{(6)}_{\bm{1}}[0:2] + C^{(6)}_{\bm{1}} [2:2] x^2 + c^2 \bigl(
            C^{(4)}_{\bm{r}}[0:1] + C^{(4)}_{\bm{r}}[2:2] x^2
        \bigr)
    \Bigr],
    \\
    \mc{Y}_3(c) &= 3 \bigl(
        C^{(6)}_{\bm{1}} [0:3] + 
        c^2 C^{(4)}_{\bm{r}}[0:3]
    \bigr),
    \\
    \mc{Y}_4(c) &= 4 \bigl(
        C^{(6)}_{\bm{1}}[0:4] + c^2 C^{(4)}_{\bm{r}}[0:4]
    \bigr).
    \label{eq:mcY4-app}
\end{align}
Here, we assume the expansion of fluctuations up to the fourth order as discussed in the previous appendix, then $\mc{X}_{4}$, $\mc{Y}_3$, and $\mc{Y}_{4}$ do not have the $x$ and $y$ dependence.
The explicit forms are summarized Tabs.~\ref{tab:potential_i_expand4_xdiff_coefficient-app} and \ref{tab:potential_i_expand4_ydiff_coefficient-app}.

\begin{table}[t]
    \centering
    \begin{tabular}{c c c}\hline 
        {} & $V_{6+4}$ & $V_{6+4'}$ 
        \\ \hhline{===} 
        $\mathcal{X}_2$ &
        $\begin{matrix}
            2836.9 -2836.9y-7092.2y^2~~~~ \\ \qquad - c^2 (304.57 - 845.88 y + 1101.5 y^2)
        \end{matrix}$   & 
        $\begin{matrix}
            2836.9 -2836.9y-7092.2y^2~~~~ \\ \qquad + c^2 (20.224+128.49 y+360.09 y^2)
        \end{matrix}$
        \\[+10pt]
        $\mathcal{X}_4$ & $-15532+1197.0 c^2$ & $-15532-62.735 c^2$
        \\ \hline
    \end{tabular}
    \caption{The concrete forms of the coefficients in Eqs.~\eqref{eq:mcX2-app} and \eqref{eq:mcX4-app}.}
    \label{tab:potential_i_expand4_xdiff_coefficient-app}
\end{table}

\begin{table}[t]
    \centering
    \begin{tabular}{cc c}\hline 
        {} & $V_{6+4}$ & $V_{6+4'}$ \\ 
        \hhline{===}
        $\mathcal{Y}_1$ & $(-2836.9+845.88 c^2) x^2$ & $(-2836.9+128.49 c^2) x^2$ 
        \\[+5pt]
        $\mathcal{Y}_2$ & 
        $\begin{matrix} 2(2836.9-7092.2 x^2)~~~~ \\ \qquad +2c^2 \left(236.75 -1101.5 x^2\right) \end{matrix}$ & 
        $\begin{matrix} 2(2836.9-7092.2 x^2)~~~~ \\ \qquad -2c^2 \left(88.039-360.09 x^2\right) \end{matrix}$ 
        \\[+10pt]
        $\mathcal{Y}_3$ & $3 \left(2836.9+236.75 c^2\right)$ & $3 \left(2836.9-88.039c^2\right)$ 
        \\
        $\mathcal{Y}_4$ & $4 \left(8439.7+621.76 c^2\right)$ & $4 \left(8439.7+11.619 c^2\right)$ 
        \\ \hline
    \end{tabular}
    \caption{The concrete forms of the coefficients in Eqs.~\eqref{eq:mcY1-app}--\eqref{eq:mcY4-app}.}
    \label{tab:potential_i_expand4_ydiff_coefficient-app}
\end{table}

\subsection{Potential expansion around $\tau = \omega$}

As similar to Eq.~\eqref{eq:Vkrexpansion-aroundi}, we introduce the expansion coefficients around $\tau = \omega$ as
\begin{align}
	V^{(k)}_{\bm{r}} = \sum_{i,j} D^{(k)}_{\bm{r}}[i:j] x^i y^j
\end{align}
where we introduce the fluctuation by $\tau = \omega + x + \iu y$.
The concrete relations between these coefficients and the modular forms are given below.

\paragraph{Weight 4 trival singlet.}

For the weight 4 trivial singlet, the coefficients are 
\begin{subequations}
\begin{align}
	D^{(4)}_{\bm{1}}[0:2]&= 3 \left| \frac{\dd \hat{Y}^{(4)}_{\bm{1}}}{\dd u} \right|^2,
	\\
	D^{(4)}_{\bm{1}}[0:3]&=- 2 \sqrt{3} \left| \frac{\dd \hat{Y}^{(4)}_{\bm{1}}}{\dd u} \right|^2,
	\\
	D^{(4)}_{\bm{1}}[0:4]&= - \left| \frac{\dd \hat{Y}^{(4)}_{\bm{1}}}{\dd u} \right|^2,
	\\
	D^{(4)}_{\bm{1}}[2:0] &= 3 \left| \frac{\dd \hat{Y}^{(4)}_{\bm{1}}}{\dd u} \right|^2,
	\\
	D^{(4)}_{\bm{1}}[2:1]&=- 2\sqrt{3} \left| \frac{\dd \hat{Y}^{(4)}_{\bm{1}}}{\dd u} \right|^2,
	\\
	D^{(4)}_{\bm{1}}[2:2]&= - 6 \left| \frac{\dd \hat{Y}^{(4)}_{\bm{1}}}{\dd u} \right|^2,
	\\
	D^{(4)}_{\bm{1}}[4:0]&=-5 \left| \frac{\dd \hat{Y}^{(4)}_{\bm{1}}}{\dd u} \right|^2
	\\
	D^{(4)}_{\bm{1}}[0:0] &= D^{(4)}_{\bm{1}}[0:1] = D^{(4)}_{\bm{1}}[1:0] = D^{(4)}_{\bm{1}}[1:1] = D^{(4)}_{\bm{1}}[1:2] = D^{(4)}_{\bm{1}}[1:3] 
	\nn \\
	& = D^{(4)}_{\bm{1}} [3:0] = D^{(4)}_{\bm{1}}[3:1] =0,
\end{align}
\label{eq:coefficients-V41-tauomega}
\end{subequations}
and their numerical values are listed in Tab.~\ref{tab:table-D4bm1}.
The potential expansion is expressed as 
\begin{align}
	   V^{(4)}_{\bm{1}}(\omega+x+ \iu  y)
        \approx&
        328y^2-379 y^3-109 y^4
        +\left(328-379y-656y^2\right)\,x^2 
        -546x^4.
\end{align}
\begin{table}
	\centering
	\begin{tabular}{|c||c|c|c|c|c|}\hline
		$i \backslash j$ & $0$ & $1$ & $2$ & $3$ & $4$ \\
		\hhline{|=#=|=|=|=|=|}
			$0$ & $0$ & $0$ & $327.80$ & $-378.51$ & $-109.27$
			\\ \hline
			$1$ & $0$ & $0$ & $0$ & $0$ & --
			\\ \hline
			$2$ & $327.80$ & $-378.51$ & $-655.60$ & -- & --
			\\ \hline
			$3$ & $0$ & $0$ & -- & -- & --
			\\ \hline
			$4$ & $-546.33$ & -- & -- & -- & --
			\\ \hline
	\end{tabular}
	\caption{
		Numerical values of $D^{(4)}_{\bm{1}}[i:j]$.
	}
	\label{tab:table-D4bm1}
\end{table}

\paragraph{Weight 4 non-trivial singlet.}

For the weight 4 non-trivial singlet, the expansion coefficients are given by 
\begin{subequations}
\begin{align}
	D^{(4)}_{\bm{1}'}[0:0] &=9 |\hat{Y}^{(4)}_{\bm{1}'}|^2,
	\\
	D^{(4)}_{\bm{1}'}[0:2] &= -12 |\hat{Y}^{(4)}_{\bm{1}'}|^2
	\\
	D^{(4)}_{\bm{1}'}[0:3] &=\frac{1}{2 \sqrt{3}} \biggl[
		\ol{\frac{\dd^3 \hat{Y}^{(4)}_{\bm{1}'}}{\dd u^3}} \hat{Y}^{(4)}_{\bm{1}'} + \frac{\dd^3 \hat{Y}^{(4)}_{\bm{1}'}}{\dd u^3} \ol{\hat{Y}^{(4)}_{\bm{1}'}} + 48 |\hat{Y}^{(4)}_{\bm{1}'}|^2
	\biggr]
	\\
	D^{(4)}_{\bm{1}'}[0:4] &=
		- \frac{1}{2}
        \ol{\frac{\dd^3 \hat{Y}^{(4)}_{\bm{1}'}}{\dd u^3}} \hat{Y}^{(4)}_{\bm{1}'}
        - \frac{1}{2}
        \frac{\dd^3 \hat{Y}^{(4)}_{\bm{1}'}}{\dd u^3} \ol{\hat{Y}^{(4)}_{\bm{1}'}}
        - 6 |\hat{Y}^{(4)}_{\bm{1}'}|^2,
	\\
	D^{(4)}_{\bm{1}'}[1:2] &= \frac{\sqrt{3}\iu}{2} \biggl[
		\ol{\frac{\dd^3 \hat{Y}^{(4)}_{\bm{1}'}}{\dd u^3}} \hat{Y}^{(4)}_{\bm{1}'} - \frac{\dd^3 \hat{Y}^{(4)}_{\bm{1}'}}{\dd u^3} \ol{\hat{Y}^{(4)}_{\bm{1}'}}
	\biggr]
	\\
	D^{(4)}_{\bm{1}'}[1:3] &=2 \iu \biggl[
		- \ol{\frac{\dd^3 \hat{Y}^{(4)}_{\bm{1}'}}{\dd u^3}} \hat{Y}^{(4)}_{\bm{1}'} + \frac{\dd^3 \hat{Y}^{(4)}_{\bm{1}'}}{\dd u^3} \ol{\hat{Y}^{(4)}_{\bm{1}'}}
	\biggr]
	\\
	D^{(4)}_{\bm{1}'}[2:0] &= - 12 |\hat{Y}^{(4)}_{\bm{1}'}|^2
	\\
	D^{(4)}_{\bm{1}'}[2:1] &=- \frac{\sqrt{3}}{2} \biggl[
		\ol{\frac{\dd^3 \hat{Y}^{(4)}_{\bm{1}'}}{\dd u^3}} \hat{Y}^{(4)}_{\bm{1}'} + \frac{\dd^3 \hat{Y}^{(4)}_{\bm{1}'}}{\dd u^3} \ol{\hat{Y}^{(4)}_{\bm{1}'}} -16 |\hat{Y}^{(4)}_{\bm{1}'}|^2
	\biggr]
	\\
	D^{(4)}_{\bm{1}'}[2:2] &=3 \ol{\frac{\dd^3 \hat{Y}^{(4)}_{\bm{1}'}}{\dd u^3}} \hat{Y}^{(4)}_{\bm{1}'} + 3 \frac{\dd^3 \hat{Y}^{(4)}_{\bm{1}'}}{\dd u^3} \ol{\hat{Y}^{(4)}_{\bm{1}'}} + 4 |\hat{Y}^{(4)}_{\bm{1}'}|^2
	\\
	D^{(4)}_{\bm{1}'}[3:0] &=\frac{\iu}{2 \sqrt{3}} \biggl[
		- \ol{\frac{\dd^3 \hat{Y}^{(4)}_{\bm{1}'}}{\dd u^3}} \hat{Y}^{(4)}_{\bm{1}'} + \frac{\dd^3 \hat{Y}^{(4)}_{\bm{1}'}}{\dd u^3} \ol{\hat{Y}^{(4)}_{\bm{1}'}}
	\biggr]
	\\
	D^{(4)}_{\bm{1}'}[3:1] &=2\iu \biggl[
	\ol{\frac{\dd^3 \hat{Y}^{(4)}_{\bm{1}'}}{\dd u^3}} \hat{Y}^{(4)}_{\bm{1}'} - \frac{\dd^3 \hat{Y}^{(4)}_{\bm{1}'}}{\dd u^3} \ol{\hat{Y}^{(4)}_{\bm{1}'}}
	\biggr]
	\\
	D^{(4)}_{\bm{1}'}[4:0] &= 
		- \frac{1}{2}
        \ol{\frac{\dd^3 \hat{Y}^{(4)}_{\bm{1}'}}{\dd u^3}} \hat{Y}^{(4)}_{\bm{1}'} 
        - \frac{1}{2}
        \frac{\dd^3 \hat{Y}^{(4)}_{\bm{1}'}}{\dd u^3} \ol{\hat{Y}^{(4)}_{\bm{1}'}} 
        + 10 |\hat{Y}^{(4)}_{\bm{1}'}|^2,
	\\
	D^{(4)}_{\bm{1}'}[0:1] &=D^{(4)}_{\bm{1}'}[1:0] = D^{(4)}_{\bm{1}'}[1:1] 
	=0,
\end{align}
\label{eq:coefficients-V41p-tauomega}
\end{subequations}
and the numerical values are summarized in Tab.~\ref{tab:table-D4bm1p}.
\begin{align}
	V^{(4)}_{\bm{1}'}(\omega + x + \iu y)
    &\approx 
    36.9 - 49.2y^2 - 351y^3 + 485y^4 \notag
    \\ &\quad
    + (
        -49.2 + 940y + 3.04 \times 10^3 y^2
    )x^2
    + 551x^4
\end{align}
\begin{table}
	\centering
	\begin{tabular}{|c||c|c|c|c|c|}\hline
		$i \backslash j$ & $0$ & $1$ & $2$ & $3$ & $4$ \\
		\hhline{|=#=|=|=|=|=|}
			$0$ & $36.904$ & $0$ & $-49.206$ & $-351.30$ & $485.45$
			\\ \hline
			$1$ & $0$ & $0$ & $0$ & $0$ & --
			\\ \hline
			$2$ & $-49.206$ & $940.26$ & $3043.9$ & -- & --
			\\ \hline
			$3$ & $0$ & $0$ & -- & -- & --
			\\ \hline
			$4$ & $551.06$ & -- & -- & -- & --
			\\ \hline
	\end{tabular}
	\caption{
		Numerical values of $D^{(4)}_{\bm{1}'}[i:j]$.
	}
	\label{tab:table-D4bm1p}
\end{table}

\paragraph{Weight 6 trivial singlet.}

The expansion coefficients for the weight 6 trivial singlet become
\begin{subequations}
\begin{align}
	D^{(6)}_{\bm{1}}[0:0] &= 27 | \hat{Y}^{(6)}_{\bm{1}} |^2,
	\\
	D^{(6)}_{\bm{1}}[0:2] &= - 54 |\hat{Y}^{(6)}_{\bm{1}}|^2,
	\\
	D^{(6)}_{\bm{1}}[0:3] &= 
        \frac{\sqrt{3}}{2} \biggl[
		\ol{\frac{\dd^3 \hat{Y}^{(6)}_{\bm{1}}}{\dd u^3}} \hat{Y}^{(6)}_{\bm{1}} 
        + \frac{\dd^3 \hat{Y}^{(6)}_{\bm{1}}}{\dd u^3} \ol{\hat{Y}^{(6)}_{\bm{1}}} 
	    \biggr]
        + 36\sqrt{3} |\hat{Y}^{(6)}_{\bm{1}}|^2,
	\\
	D^{(6)}_{\bm{1}}[0:4] &= 
        - \frac{3}{2} \biggl[
		\ol{\frac{\dd^3 \hat{Y}^{(6)}_{\bm{1}}}{\dd u^3}} \hat{Y}^{(6)}_{\bm{1}} 
        + \frac{\dd^3 \hat{Y}^{(6)}_{\bm{1}}}{\dd u^3} \ol{\hat{Y}^{(6)}_{\bm{1}}} 
	    \biggr]
        -9 | \hat{Y}^{(6)}_{\bm{1}}|^2,
	\\
	D^{(6)}_{\bm{1}}[1:2] & = 
        \frac{3\sqrt{3}\iu}{2} \biggl[
		\ol{\frac{\dd^3 \hat{Y}^{(6)}_{\bm{1}}}{\dd u^3}} \hat{Y}^{(6)}_{\bm{1}} 
        - \frac{\dd^3 \hat{Y}^{(6)}_{\bm{1}}}{\dd u^3} \ol{\hat{Y}^{(6)}_{\bm{1}}}
	\biggr],
	\\
	D^{(6)}_{\bm{1}}[1:3] &= 
        -6 \iu \biggl[
		\ol{\frac{\dd^3 \hat{Y}^{(6)}_{\bm{1}}}{\dd u^3}} \hat{Y}^{(6)}_{\bm{1}} 
        - \frac{\dd^3 \hat{Y}^{(6)}_{\bm{1}}}{\dd u^3} \ol{\hat{Y}^{(6)}_{\bm{1}}}
	\biggr],
	\\
	D^{(6)}_{\bm{1}}[2:0] &= -54 |\hat{Y}^{(6)}_{\bm{1}}|^2,
	\\
	D^{(6)}_{\bm{1}}[2:1] &= 
        - \frac{3 \sqrt{3}}{2} \biggl[
		\ol{\frac{\dd^3 \hat{Y}^{(6)}_{\bm{1}}}{\dd u^3}} \hat{Y}^{(6)}_{\bm{1}} 
        + \frac{\dd^3 \hat{Y}^{(6)}_{\bm{1}}}{\dd u^3} \ol{\hat{Y}^{(6)}_{\bm{1}}} 
	   \biggr]
        + 36\sqrt{3} |\hat{Y}^{(6)}_{\bm{1}}|^2,
	\\
	D^{(6)}_{\bm{1}}[2:2]&= 
        9 \biggl[
		\ol{\frac{\dd^3 \hat{Y}^{(6)}_{\bm{1}}}{\dd u^3}} \hat{Y}^{(6)}_{\bm{1}} 
        + \frac{\dd^3 \hat{Y}^{(6)}_{\bm{1}}}{\dd u^3} \ol{\hat{Y}^{(6)}_{\bm{1}}} 
	   \biggr]
        + 6 |\hat{Y}^{(6)}_{\bm{1}}|^2,
	\\
	D^{(6)}_{\bm{1}}[3:0] &= - \frac{\sqrt{3}\iu}{2} \biggl[
		\ol{\frac{\dd^3 \hat{Y}^{(6)}_{\bm{1}}}{\dd u^3}} \hat{Y}^{(6)}_{\bm{1}} - \frac{\dd^3 \hat{Y}^{(6)}_{\bm{1}}}{\dd u^3} \ol{\hat{Y}^{(6)}_{\bm{1}}}
	\biggr],
	\\
	D^{(6)}_{\bm{1}}[3:1] &= 6 \iu \biggl[
		\ol{\frac{\dd^3 \hat{Y}^{(6)}_{\bm{1}}}{\dd u^3}} \hat{Y}^{(6)}_{\bm{1}} - \frac{\dd^3 \hat{Y}^{(6)}_{\bm{1}}}{\dd u^3} \ol{\hat{Y}^{(6)}_{\bm{1}}}
	\biggr],
	\\
	D^{(6)}_{\bm{1}}[4:0] &= 
        - \frac{3}{2} \biggl[
		\ol{\frac{\dd^3 \hat{Y}^{(6)}_{\bm{1}}}{\dd u^3}} \hat{Y}^{(6)}_{\bm{1}} 
        + \frac{\dd^3 \hat{Y}^{(6)}_{\bm{1}}}{\dd u^3} \ol{\hat{Y}^{(6)}_{\bm{1}}} 
	    \biggr]
        + 63 |\hat{Y}^{(6)}_{\bm{1}}|^2,
	\\
	D^{(6)}_{\bm{1}}[0:1] &= D^{(6)}_{\bm{1}}[1:0] = D^{(6)}_{\bm{1}}[1:1] =0,
\end{align}
\label{eq:coefficients-V61-tauomega}
\end{subequations}
and their numerical values are listed in Tab.~\ref{tab:table-D6bm1}.
We expand the potential as 
\begin{align}
	V^{(6)}_{\bm{1}}(\omega+x+ \iu y)
    &=
    224 -448 y^2+ 7510 y^3 - 12200 y^4
	\nn \\
    &
    \quad 
    +\left(-449 -20500y +73200y^2\right)x^2  
    -11600x^4.
\end{align}

\begin{table}
	\centering
	\begin{tabular}{|c||c|c|c|c|c|}\hline
		$i \backslash j$ & $0$ & $1$ & $2$ & $3$ & $4$ \\
		\hhline{|=#=|=|=|=|=|}
			$0$ & $224.19$ & $0$ & $-448.38$ & $7514.8$ & $-12194$
			\\ \hline
			$1$ & $0$ & $0$ & $0$ & $0$ & --
			\\ \hline
			$2$ & $-448.38$ & $-20473$ & $73164$ & -- & --
			\\ \hline
			$3$ & $0$ & $0$ & -- & -- & --
			\\ \hline
			$4$ & $-11596$ & -- & -- & -- & --
			\\ \hline
	\end{tabular}
	\caption{
		Numerical values of $D^{(6)}_{\bm{1}}[i:j]$.
	}
	\label{tab:table-D6bm1}
\end{table}

\section{Explicit form of K\"{a}hler metric}
\label{sec:Kahler-metric}

\subsection{Single stabilizer case}

When the number of the stabilizer field is unity, we assume the following K\"{a}hler potential
\begin{align}
	\mc{K} = Z_X|X|^2 - \frac{Z_X^2 |X|^4}{4 \Lambda_X^2} - h \log(- \iu \tau + \iu \bar{\tau}),
	\qquad 
	Z_X = (- \iu \tau + \iu \bar{\tau})^{k_X}.
\end{align}
The K\"{a}hler metric is introduced by 
\begin{align}
	\mc{K}_{i\bar{j}} &= \begin{pmatrix}
		\mc{K}_{\tau \bar{\tau}} & \mc{K}_{\tau\bar{X}} \\
		\mc{K}_{X \bar{\tau}} & \mc{K}_{X \bar{X}}
	\end{pmatrix},
	\label{eq:Kahler-metric-single-modular-form}
\end{align}
and the explicit forms of these components are given by 
\begin{subequations}
\begin{align}
	\mc{K}_{\tau \bar{\tau}} &= \frac{2 h  +2 (-1 + k_X) k_X(2 \Im \tau)^{k_X} |X|^2- k_X (-1 + 2k_X)k_X (2 \Im \tau)^{2 k_X}|X|^4/\Lambda_X^2}{8 (\Im \tau)^2},
	\\
	\mc{K}_{\tau \bar{X}} &= - \iu k_X (2 \Im \tau)^{-1 + k_X} X(1 - (2 \Im \tau)^{k_X} |X|^2/\Lambda_X^2 ,
	\\
	\mc{K}_{X \bar{\tau}} &= \iu k_X (2 \Im \tau)^{-1 + k_X} \bar{X} (1 - (2 \Im \tau)^{k_X} |X|^2/\Lambda_X^2,
	\\
	\mc{K}_{X \bar{X}} &= (2 \Im \tau)^{k_X}(1 - (2 \Im \tau)^{k_X} |X|^2 /\Lambda_X^2).
	\label{eq:Kahler-metric-single-modular-form2}
\end{align}
\end{subequations}
If the VEV of $X$ is $\braket{X} = 0$, $\mc{K}_{X\bar{X}}$ is reduced to $Z_X$.

\subsection{Double stabilizer case}

The K\"{a}hler potential is
\begin{align}
	\mc{K} &= Z_X |X|^2 - \frac{Z_X^2 |X|^4}{4\Lambda_X^2} + Z_x |x|^2 - \frac{Z_\Xi |\Xi|^4}{4 \Lambda_\Xi^2} - h \log (-\iu \tau + \iu \bar{\tau}),
	\\
	Z_X &= (-\iu \tau + \iu \bar{\tau})^{k_X},
	\qquad 
	Z_\Xi =(-\iu \tau + \iu \bar{\tau})^{k_\Xi}.
\end{align}
We write the K\"{a}hler metric of this case as 
\begin{align}
	\mc{K}_{i\bar{j}} &= \begin{pmatrix}
		\mc{K}_{\tau \bar{\tau}} & \mc{K}_{\tau\bar{X}} & \mc{K}_{\tau \bar{\Xi}} \\
		\mc{K}_{X \bar{\tau}} & \mc{K}_{X \bar{X}} & \mc{K}_{X \bar{\Xi}} \\
		\mc{K}_{\Xi \bar{\tau}} & \mc{K}_{\Xi \bar{X}} & \mc{K}_{\Xi \bar{\Xi}}
	\end{pmatrix},
\end{align}
and the components are 
\begin{align}
	\mc{K}_{\tau \bar{\tau}} &= \frac{1}{8 (\Im \tau)^2} \biggl[2 h  + 2 (-1 + k_X) k_X(2 \Im \tau)^{k_X} |X|^2- k_X (-1 + 2k_X)k_X (2 \Im \tau)^{2 k_X}|X|^4/\Lambda_X^2 
	\nn \\
	& \qquad + 2 (-1 + k_\Xi) k_\Xi(2 \Im \tau)^{k_\Xi} |\Xi|^2- k_\Xi (-1 + 2k_\Xi)k_\Xi (2 \Im \tau)^{2 k_\Xi}|\Xi|^4/\Lambda_\Xi^2
	\biggr],
	\\
	\mc{K}_{\tau \bar{X}} &= - \iu k_X (2 \Im \tau)^{-1 + k_X} X(1 - (2 \Im \tau)^{k_X} |X|^2/\Lambda_X^2 ),
	\\
	\mc{K}_{\tau \bar{\Xi}} &=- \iu k_\Xi (2 \Im \tau)^{-1 + k_\Xi} \Xi (1 - (2 \Im \tau)^{k_\Xi} |\Xi|^2/\Lambda_\Xi^2),
	\\
	\mc{K}_{X \bar{\tau}} &= \iu k_X (2 \Im \tau)^{-1 + k_X} \bar{X} (1 - (2 \Im \tau)^{k_X} |X|^2/\Lambda_X^2),
	\\
	\mc{K}_{X\bar{X}} &= (2 \Im \tau)^{k_X}(1 - (2 \Im \tau)^{k_X} |X|^2 /\Lambda_X^2),
	\\
	\mc{K}_{\Xi \bar{\tau}} &= \iu k_\Xi (2 \Im \tau)^{-1 + k_\Xi} \bar{\Xi} (1 - (2 \Im \tau)^{k_\Xi} |\Xi|^2/\Lambda_\Xi^2),
	\\
	\mc{K}_{\Xi\bar{\Xi}} &= (2 \Im \tau)^{k_\Xi}(1 - (2 \Im \tau)^{k_\Xi} |\Xi|^2 /\Lambda_\Xi^2),
	\\
	\mc{K}_{X\bar{\Xi}} &= \mc{K}_{\Xi\bar{X}} = 0.
\end{align}
We note that $\mc{K}_{X\bar{\Xi}}$ and $\mc{K}_{\Xi\bar{X}}$ are vanished because of $\rU(1)_R$ symmetry.
If $\braket{X} = \braket{\Xi} = 0$, the diagonal components reduce as $\mc{K}_{X\bar{X}} = Z_X$, $\mc{K}_{\Xi \bar{\Xi}} = Z_\Xi$.

\section{Hessian analysis}
\label{sec:Hessian-analysis}

In this appendix, we will summarize the components of the Hessian matrices in the potential vacua for $V_{6+4}$, $V_{6+4'}$, and $V_{4+4'}$, respectively.
The definition is given by Eq.~\eqref{eq:hessian-matrix}.
Here we leave the fluctuation dependence of $(x,y)$, which is introduced by $\tau = \tau_* + x + \iu y$, where $\tau_*$ denotes the modulus VEV.

\subsection{$V_{6+4}$}

\begin{itemize}
	\item $\tau = \iu$
	\begin{align}
		H_{11}(x,y;c) &= 10^3(-186 x^2 + 2 (2.84 - 2.84 y - 7.09 y^2)) 
		\nn \\
		& \qquad 
		+ c^2 ( 14400 x^2 + 2 (-305 + 846 y - 1100 y^2)),
		\\
		H_{12} (x,y;c) &= H_{21}(x,y;c) = 2c^2 x (846-2200y),
		\\
		H_{22}(x,y;c) &= 10^3(5.68 - 14.18 x^2 - 17.04 y + 101.28 y^2) 
		\nn \\
		& \qquad 
		+ c^2(474 - 2200 x^2 - 1422 y + 7464 y^2).
	\end{align}
	\item $\tau = \omega$
	\begin{align}
		H_{11}(x,y;c) &= -139200 x^2 + 2(-449 - 20500 y + 73200 y^2) 
		\nn \\
		& \qquad 
		+ c^2 ( -6552 x^2 + 2 (328 - 379 y - 656 y^2)),
		\\
		H_{12} (x,y;c) &= H_{21}(x,y;c) = 2 c^2 x (-379-1310y),
		\\
		H_{22}(x,y;c) &= -896 + 146 400 x^2 + 45060 y- 146400 y^2
		\nn \\
		& \qquad 
		+ c^2 ( 656 - 1312 x^2 - 2274 y - 1308 y^2).
	\end{align}
\end{itemize}

\subsection{$V_{6+4'}$}

\begin{itemize}
	\item $\tau = \iu$
	\begin{align}
    		H_{11}(x,y;c) & \sim 5.67\times10^3 + c^2 (720 y^2 + 257 y + 40.4)
    		\nn \\
    		& \qquad 
    		- (1.86\times 10^5 + 753 c^2)x^2,
    		\\
    		H_{12}(x,y;c) & =H_{21}(c,x,y) 
            \sim1.44\times10^3  c^2 xy + 257 c^2 x,
    		\\
    		H_{22}(x,y;c) & \sim 
            5.67\times10^3 - 176 c^2 
            - (1.42 \times 10^4 - 720 c^2)x^2
    		\nn \\
    		& \qquad 
    		+ (1.70 \times 10^4 - 528 c^2) y
    		+ (1.01 \times 10^5 + 139 c^2 )y^2.
	\end{align}
	\item $\tau = \iu \infty$
	\begin{align}
    		H_{11}(x,y,;c) &= (2.5805 \times 10^5) \pi^2 y^6 \ee^{-2 \pi y} \cos ( 2\pi x),
    		\label{eq:H11-V64p-iinf}
    		\\
    		H_{12} (x,y,;c) &= H_{21}(x,y,;c) = - (2.5805 \times 10^{5}) \pi y^5 ( \pi y - 3) \ee^{-2\pi y} \sin (2 \pi x),
    		\\
    		H_{22}(x,y,;c) &= 
            128 y^2 \bigg\{
            3 y^2 
            \left(84672 e^{-4 \pi  y} 
            \left(8 \pi ^2 y^2-24 \pi  y+15\right)+5\right)
            \nn \\ & \qquad
            -
            1008 e^{-2 \pi  y} y^2
            \left(15 - 12 \pi y + 2 \pi ^2 y^2 \right) 
            \cos (2 \pi  x)
            \nn \\ & \qquad
            +
            c^2 8 \mr{e}^{-\frac{4 \pi  y}{3}} 
            \left( 4 \pi ^2 y^2-24 \pi  y+27 \right)
        \bigg\}.
	\end{align}

\end{itemize}

\subsection{$V_{4+4'}$}

\begin{itemize}
	\item $\tau = \omega$
	\begin{align}
		H_{11}(x,y;c) &=
            655.60-6556.0 x^2
            \nn \\
            &\qquad
            +c^2 \left(6612.7 x^2+6087.8 y^2+1880.5 y-98.411\right),
		\\
        H_{12}(x,y;c) &=
            2 c^2 (940.26+6087.8 y)x,
        \\
		H_{22}(x,y;c) &= 
            655.60+2271.0 y-1311.2 y^2-1311.2 x^2
            \nn \\
            & \qquad
            +c^2 \left(-98.411+ 2107.8 y +5825.4 y^2 +6087.8 x^2\right).
	\end{align}
	\item $\tau = -1/2 + \iu \infty $
	\begin{align}
		H_{11} (x,y;c)&=
            -30720 \pi^2y^4\mr{e}^{-2\pi y}\cos(2\pi x),
		\\
		H_{12}(x,y;c) &= H_{21}(x,y,;c) 
            = 30720 \pi  (\pi y-2) y^3e^{-2 \pi  y} \sin (2 \pi  x),
		\\
		H_{22}(x,y;c) &= 
            64 y^2 \bigg\{
                3
                +
                480 e^{-2 \pi  y} 
                \left(3-4 \pi  y+\pi ^2 y^2\right) \cos (2 \pi  x)
                \nn \\ & \qquad \qquad
                +
                57600 e^{-4 \pi  y}
                \left(3-8 \pi  y+4 \pi ^2 y^2\right)
                \nn \\ & \qquad \qquad
                +
                16 c^2 e^{-\frac{4 \pi  y}{3}} 
                \left(27-24 \pi  y+4 \pi ^2 y^2\right)
            \bigg\}.
	\end{align}
\end{itemize}

{\small
\newcommand{\arxivfont}{\rmfamily}
\bibliographystyle{yautphysm}
\bibliography{reff}
}

\end{document}